\documentclass[twocolumn,preprintnumbers]{revtex4}
\usepackage{epsf}

\begin{document}
\preprint{KUNS-2457}
\title{Cluster states and monopole transitions in $^{16}$O}
\author{Yoshiko Kanada-En'yo}
\affiliation{Department of Physics, Kyoto University, Kyoto 606-8502, Japan}
\begin{abstract}
Cluster structures and monopole transitions in positive parity states of $^{16}$O were 
investigated based on the generator coordinate method calculation of an extended 
$^{12}$C+$\alpha$ cluster model. 
The ground and excited states of a $^{12}$C cluster are taken into account by using 
$^{12}$C wave functions obtained with the method of antisymmetrized molecular dynamics. 
The $0^+_2$ state of $^{16}$O
and its rotational members, 
the $2^+_1$ and $4^+_1$ states are described well by the cluster states dominated by 
the $^{12}$C($0^+_1$)+$\alpha$ structure. 
Above the $^{12}$C($0^+_2$)+$\alpha$ threshold energy, we obtained  
a $0^+$ state having the 
$^{12}$C($0^+_2$)+$\alpha$ cluster structure, which is considered to be a candidate for
the $4\alpha$ cluster gas state.
The band structures were discussed 
based on the calculated $E2$ transition strength.
Isoscalar Monopole excitations from the ground state were also discussed.
\end{abstract}


\maketitle

\section{Introduction}
Cluster structure is one of the essential features of nuclei as well as mean-field feature. 
Well developed cluster structures have been known, in particular, 
in excited states of stable light nuclei and also discovered in unstable nuclei. 
In these years, a new type of cluster state, a $\alpha$ cluster 
gas state, has been suggested in light $Z=N$ nuclei \cite{Tohsaki:2001an,Funaki:2002fn,Yamada:2003cz,Funaki:2003af,Funaki:2008gb,Funaki:2010px,Yamada:2011bi}. 
It has been proposed that 2$\alpha$ and 3$\alpha$ cluster gas states are
realized in the $0^+_1$ state of $^8$Be and
the $0^+_2$ state of $^{12}$C, where all $\alpha$ clusters are almost freely moving in a dilute
density like a gas. It is a challenging problem to search for such cluster gas states in other nuclei. 
For instance, possibility of 
$\alpha$ cluster gas states in $Z=N=2n$ nuclei up to $^{40}$Ca was discussed
in the systematic study with a non-microscopic cluster model, which suggested that 
$\alpha$ cluster gases may appear in the energy region near the corresponding $n$-$\alpha$ break-up 
threshold consistently to the Ikeda threshold rule\cite{Yamada:2003cz}. 
Cluster gas states including non-$\alpha$ clusters or those around a core nucleus
were also suggested in excited states of $^{11}$B, $^8$He and $^{10}$Be\cite{KanadaEn'yo:2006bd,KanadaEn'yo:2007ie,Itagaki:2007rv,Suhara:2012zr,Ichikawa:2012mk,Kobayashi:2012di,Kobayashi:2013iwa}.

Recently, the search for $4\alpha$ cluster gas state in excited states of $^{16}$O has been performed
in experimental and theoretical works\cite{Funaki:2008gb,Funaki:2010px,Wakasa:2007zza}. 
The semi-microscopic 4$\alpha$ calculation 
by Funaki {\it et al.} suggested that the $^{16}$O($0^+_6$) state near 
the $4\alpha$ threshold has the large 
$^{12}$C$(0^+_2)+\alpha$ component and is a candidate for the dilute 4$\alpha$ cluster gas state 
\cite{Funaki:2008gb,Funaki:2010px}.
It is also an interesting problem to assign band members of the cluster gas state to clarify the
property of the cluster gas, especially, stability against rotation as discussed 
in Refs.~\cite{Ohkubo:2010zz,Funaki2011-ykis}.

$^{16}$O is a double closed-shell nuclei and its ground state is dominated by $p$-shell closed
configuration, while there exist many excited states that 
are difficult to be described by a simple shell model.
Semi-microscopic and microscopic $^{12}$C+$\alpha$ cluster 
models \cite{suzuki76,Libert80,supp80} were applied to study excited 
states of $^{16}$O and it has been shown that 
many excited states can be described by $^{12}$C+$\alpha$ cluster structures. 
For instance, in the calculation with the 
$^{12}$C+$\alpha$ orthogonality condition model (OCM) \cite{suzuki76}, 
a semi-microscopic cluster model \cite{OCM},  
the $0^+_2$ state of $^{16}$O and its rotational band members, the $2^+_1$ and $4^+_1$
 states,
are described by the cluster state having the dominant $^{12}$C($0^+_1$)+$\alpha$ component.
Moreover, the $^{16}$O($0^+_3$) state is considered to mainly have the
$^{12}$C($2^+_1$)+$\alpha$ component. These results are supported also by 
4$\alpha$-OCM calculations \cite{Funaki:2008gb,Funaki:2010px,Fukatsu92}. 
Thus, many excited states up to $\sim$ 14 MeV are considered to be weak-coupling 
$^{12}$C+$\alpha$ cluster states having large components of 
$^{12}$C$(0^+_1)$+$\alpha$, $^{12}$C$(2^+_1)$+$\alpha$, and so on. 
The cluster structures of these excited states are supported by the experimental data of 
$E2$ and monopole transition strengths
as well as the $\alpha$-decay widths \cite{suzuki76,Libert80,Yamada:2011ri}.

Above these $^{12}$C+$\alpha$ cluster states, a 4$\alpha$ cluster state was predicted 
at the energy near the 4$\alpha$ and $^{12}$C($0^+_2$)+$\alpha$ threshold energies 
by Funaki {\it et al.} with the 4$\alpha$-OCM \cite{Funaki:2008gb,Funaki:2010px}. 
This state has the large $^{12}$C($0^+_2$)+$\alpha$ component, that is, 
the $3\alpha$ cluster gas state of the $^{12}$C($0^+_2$) with an additional $\alpha$ around the 
$3\alpha$ gas. The large occupation probability 
of 4 $\alpha$ particles in the same $0S$ and low-momentum orbit 
was demonstrated by the analysis of the 4$\alpha$-OCM
wave function.

In spite of the success of those calculations with the semi-microscopic cluster models such as the $^{12}$C+$\alpha$-OCM and
the $4\alpha$-OCM, there is no microscopic calculation that can 
reproduce the excitation energies of the cluster states in $^{16}$O.
The microscopic calculations with the resonating group method (RGM)\cite{RGM} and 
the generator coordinate method (GCM)\cite{GCM} 
of $^{12}$C+$\alpha$ cluster models\cite{Libert80,supp80} 
failed to reproduce the experimental excitation energy of the $0^+_2$ at 6.05 MeV. They 
largely overestimated it by a factor 2$-$3 as $E_x(0^+_2)\sim 16$ MeV.
One of the most crucial problems in microscopic calculations using effective nuclear forces for $^{16}$O
is the underbinding problem of $^{12}$C relative to $^{16}$O, or in other words, 
the overbinding problem of $^{16}$O relative to $^{12}$C. 

Our aim is to investigate cluster structures of excited states of $^{16}$O. In particular, we 
search for a highly excited $0^+$ state having the $^{12}$C($0^+_2$)+$\alpha$ structure, which may be the
candidate for the $4\alpha$ gas state.
We perform the GCM calculation of an extended $^{12}$C+$\alpha$ model. In the present calculation,
we adopt the $^{12}$C wave functions obtained with the variation after the 
parity and angular-momentum projections(VAP) 
in the framework of anstisymmetrized molecular dynamics (AMD) \cite{ENYObc,AMDsupp-rev}. 
As shown in the previous works on $^{12}$C \cite{KanadaEn'yo:1998rf,KanadaEn'yo:2006ze},
the AMD+VAP calculation succeeded to describe well the structures of ground and excited states of $^{12}$C,
such as the developed $3\alpha$-cluster structure in the 
excited states as well as the ground state properties. 
The binding energy of $^{12}$C was improved because of the energy gain of the spin-orbit force 
due to the mixing of $p_{3/2}$-shell configurations. We use the same effective nuclear force 
used in the previous study of $^{12}$C, that it, the MV1 force \cite{MVOLKOV} containing 
the phenomenological three-body 
repulsive force to avoid the overshooting problem of the binding energy in heavier nuclei.
To take into account the ground and excited states of $^{12}$C 
we superpose the $^{12}$C AMD wave functions and approximately perform the 
double projection, that is the angular-momentum projection 
of the subsystem $^{12}$C and that of the total system.
Isoscalar monopole excitations in $^{16}$O are also discussed.

This paper is organized as follows. 
In the next section, we explain the formulation of the present calculation.
The results are shown in \ref{sec:results}, and 
isoscalar monopole excitations are discussed in \ref{sec:discussions}.
Finally, a summary and outlooks are given in \ref{sec:summary}.

\section{Formulation}
\subsection{$^{12}$C(AMD)+$\alpha$GCM calculation for $^{16}$O}
The ground and excited states of $^{16}$O are described 
by using an extended $^{12}$C+$\alpha$ cluster wave function.
To describe inter-cluster motion,  
the distance $d$ between the mean positions of $^{12}$C and $\alpha$ centers
is treated as the generator coordinate, and the $^{12}$C+$\alpha$ wave functions with 
different $d$ values are superposed.
The $\alpha$ cluster is written by the $(0s)^4$ harmonic oscillator wave function
$\Phi_\alpha(3{\bf S}/4)$ which is localized around the position $3{\bf S}/4$ with ${\bf S}=(0,0,d)$.
The $^{12}$C cluster is localized around $-{\bf S}/4$ and described 
by the superposition of AMD wave functions. 

An AMD wave function for the $^{12}$C cluster 
localized around the origin 
is given as follows,
\begin{eqnarray}\label{eq:AMD}
\Phi_{^{12}{\rm C}}^{\rm AMD}({\bf Z})&=&\frac{1}{\sqrt{A_C!}} {\cal{A}_C} \{
  \varphi_1,\varphi_2,...,\varphi_{A_C} \},\\
 \varphi_i&=& \phi_{{\bf X}_i}\chi_i\tau_i,\\
 \phi_{{\bf X}_i}({\bf r}_j) & = &  \left(\frac{2\nu}{\pi}\right)^{4/3}
\exp\bigl\{-\nu({\bf r}_j-\frac{{\bf X}_i}{\sqrt{\nu}})^2\bigr\},
\label{eq:spatial}\\
 \chi_i &=& (\frac{1}{2}+\xi_i)\chi_{\uparrow}
 + (\frac{1}{2}-\xi_i)\chi_{\downarrow}.
\end{eqnarray}
Here $A_C$ is the mass number of $^{12}$C, $A_C=12$, and the operator 
${\cal{A}_C}$ is the antisymmetrizer of the $A_C$ nucleons.
The wave function $\Phi_{^{12}{\rm C}}^{\rm AMD}({\bf Z})$ is written by 
a Slater determinant of single-particle wave functions $\varphi_i$, each of 
which is given by a product of the spatial ($\phi_{{\bf X}_i}$), 
the intrinsic spin($\chi_i$), and isospin($\tau_i$) functions.
The isospin
function fixed to be up (proton) or down (neutron). 
The spatial part $\phi_{{\bf X}_i}$ is written by the Gaussian wave packet localized around 
the position ${\bf X}_i$ in the phase space.
Accordingly, an AMD wave function
is expressed by a set of variational parameters, ${\bf Z}\equiv 
\{{\bf X}_1,{\bf X}_2,\cdots, {\bf X}_{A_C},\xi_1,\xi_2,\cdots,\xi_{A_C} \}$, 
which expresses an AMD configuration of the $^{12}$C cluster.
The mean position $\{{\bf X}_1+{\bf X}_2+\cdots+{\bf X}_{A_C}\}/A_C$ of 
$^{12}$C mass center is set on the origin.

The $^{12}$C wave function is shifted from the origin to the position $-{\bf S}/4$
by shifting the Gaussian center parameters ${\bf X}_i\rightarrow {\bf X}_i-{\bf S}/4$.
The shifted $^{12}$C AMD wave function is denoted by $\Phi_{^{12}{\rm C}}^{\rm AMD}
(-{\bf S}/4;{\bf Z})$.
An wave function $\Phi_{^{12}{\rm C}}^{\rm AMD}
(-{\bf S}/4;{\bf Z})$ corresponds to the $^{12}$C cluster around $-{\bf S}/4$ having 
an intrinsic wave function specified by the set of parameters ${\bf Z}$.
To construct the 
angular-momentum eigen state of the subsystem $^{12}$C 
projected from the intrinsic state, it is necessary to
superpose rotated states of the intrinsic wave function. For an configuration 
${\bf Z}={\bf Z}^{(k)}$ ($k$ is the label for the configuration) of the 
$^{12}$C AMD wave function, we prepare rotated states $R^{\rm sub}(\Omega')\Phi_{^{12}{\rm C}}^{\rm AMD}(-{\bf S}/4;{\bf Z}^{(k)})$ of the subsystem $^{12}$C. Here  
$R^{\rm sub}(\Omega')$ is the operator of the Euler angle $\Omega'$ rotation of the 
subsystem around $-{\bf S}/4$. A wave function of $^{16}$O
is given by performing the antisymmetrization of all nucleons and 
the parity and angular-momentum projections,
\begin{eqnarray}
&&\Phi^{J\pi K}_{^{12}{\rm C}+\alpha}(d,\Omega'_j,{\bf Z}^{(k)})=\nonumber \\
&&P^{J\pi}_{MK}{\cal A}\left\{R^{\rm sub}(\Omega')
\Phi_{^{12}{\rm C}}^{\rm AMD}(-{\bf S}/4;{\bf Z}^{(k)})\cdot \Phi_\alpha(3{\bf S}/4) \right\}.
\end{eqnarray}
Here ${\cal A}$ is the antisymmetrizer for all sixteen nucleons and $P^{J\pi}_{MK}$ is 
the parity($\pi$) and angular-momentum projection operator for the total system.

We superpose $^{16}$O wave functions constructed from the $^{12}$C AMD wave function 
and the $\alpha$ cluster wave function. Each $^{16}$O wave function is specified 
by the AMD configuration ${\bf Z}^{(k)}$, the rotation angle $\Omega'_j$ for 
the $^{12}$C cluster, and the inter-cluster distance $d_i$. Then the final 
$^{16}$O wave function in the present $^{12}$C(AMD)+$\alpha$GCM model is written as follows, 
\begin{equation}
\Psi_{{\rm AMD}+\alpha{\rm GCM}}^{J^\pi_n}=\sum_{K,i,j,k}c^{J^\pi_n}(K,i,j,k)
\Phi^{J\pi K}_{^{12}{\rm C}+\alpha}(d_i,\Omega'_j,{\bf Z}^{(k)}).
\end{equation}
The coefficients $c^{J^\pi_n}(K,i,j,k)$ for the $J^\pi_n$ state are treated 
as independent parameters and they are determined by 
solving the Hill-Wheeler equation as done in the GCM\cite{GCM}. 
In principle, the superposition of rotated states of the $^{12}$C cluster 
is equivalent to the so-called "double projection", in which 
the angular-momentum projections are done for the subsystem $^{12}$C 
and also for the total system. It corresponds to take into account 
different spin states of the $^{12}$C cluster. In the practical calculation, 
however, we use only a limited number of the rotation angle $\Omega'_j$ and it is an 
approximated method of the double projection. 
By superposing several AMD configurations of $^{12}$C, excited
states as well as the ground state of the $^{12}$C cluster are incorporated.
The details of the AMD configurations of $^{12}$C are explained later. 

For general nuclei, we can consider the extended cluster model "AMD+$\alpha$GCM", 
in which a core nucleus is written by AMD wave functions and relative motion between 
an $\alpha$ cluster and the core is taken into account by superposing core-$\alpha$ cluster
wave functions with various values of the distance $d$. 
Based on a similar concept,  
core+$n$ cluster models have been already used to describe a valence neutron motion 
around the core expressed by AMD wave functions in the studies of neutron-rich nuclei.
Firstly a $^{10}$Be(AMD)+$n$GCM model without the angular-momentum projection of subsystem
has been adopted to $^{11}$Be\cite{ENYObc}, and recently, $^{30}$Ne(AMD)+$n$GCM and $^{12}$Be(AMD)+$n$GCM
models have been applied to  $^{31}$Ne and $^{13}$Be \cite{Minomo:2011bb,KanadaEn'yo:2012rm}.

\subsection{Wave functions of $^{12}$C}
In the previous work on $^{12}$C\cite{KanadaEn'yo:1998rf,KanadaEn'yo:2006ze}, 
the AMD+VAP method has been applied to $^{12}$C
and it has been proved to describe well the structures of the ground and excited 
states in $^{12}$C. To describe the $^{12}$C cluster in the 
present $^{12}$C(AMD)+$\alpha$GCM calculation, we use the intrinsic wave functions 
of $^{12}$C obtained with the AMD+VAP in Ref.~\cite{KanadaEn'yo:2006ze}. 

We here briefly explain the AMD+VAP method\cite{KanadaEn'yo:1998rf,KanadaEn'yo:2006ze}. 
More details of the method
are described in Ref.~\cite{KanadaEn'yo:2006ze}.
As mentioned before, the AMD wave function of $^{12}$C explained 
in Eq.~\ref{eq:AMD} is specified by the 
set of parameters, ${\bf Z}=
\{{\bf X}_1,{\bf X}_2,\cdots, {\bf X}_{A_C},\xi_1,\xi_2,\cdots,\xi_{A_C} \}$. 
In the AMD framework, these are treated as variational parameters and 
determined by the energy variation.
In the AMD+VAP method, the energy variation is performed 
after the spin-parity projection. Namely, 
the parameters ${\bf X}_i$ and $\xi_{i}$($i=1\sim A$) are varied to
minimize the energy expectation value of the Hamiltonian,
$\langle \Phi|H|\Phi\rangle/\langle \Phi|\Phi\rangle$,
with respect to the spin-parity eigen wave function 
$\Phi=P^{J\pi}_{MK}\Phi_{^{12}{\rm C}}^{\rm AMD}({\bf Z})$
projected 
from the AMD wave function of $^{12}$C.
Then the optimum AMD wave function
$\Phi_{^{12}{\rm C}}^{\rm AMD}({\bf Z}^{J^\pi_1})$,
which approximately describes the intrinsic wave function for 
the $J^\pi_1$ state, is obtained. For higher $J^\pi_n$ states, 
the variation is done for the component orthogonal to 
the lower $J^\pi$ states.
For each $J(k)^{\pi(k)}_{n(k)}$, the optimum parameters ${\bf Z}^{(k)}$ 
are obtained. Here $(k)$ is the label for the AMD configuration for the 
$J(k)^{\pi(k)}_{n(k)}$ state. 
After the VAP procedure, final wave functions for $J^\pi$ states are expressed 
by the superposition of the spin-parity eigen wave functions 
projected from all the intrinsic wave functions 
$\Phi^{\rm AMD}_{^{12}{\rm C}}({\bf Z}^{(k)})$ 
as, 
\begin{equation}\label{eq:diago}
\Psi^{J_n,\pi}_{^{12}{\rm C}}=\sum_{K,k}  c^{J^\pi_n}_{^{12}{\rm C}} (K,k) 
|P'^{J\pi}_{MK}\Phi_{^{12}{\rm C}}^{\rm AMD}({\bf Z}^{(k)}),
\end{equation}
where the coefficients $c^{J^\pi_n}_{^{12}{\rm C}}(K,k)$ 
are determined by solving the Hill-Wheeler equation, i.e., the 
diagonalization of the norm and Hamiltonian matrices.

\begin{figure}[th]
\epsfxsize=0.4\textwidth
\centerline{\epsffile{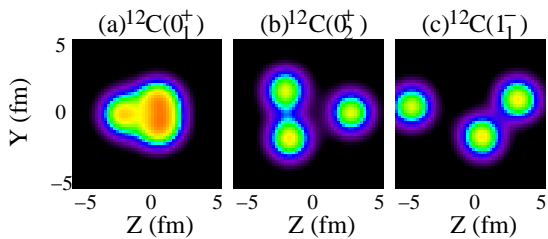}}
\caption{\label{fig:c12dense}
(Color online) 
Density distribution of the intrinsic states of (a) $^{12}$C($0^+_1$), (b) $^{12}$C($0^+_2$), and
(c) $^{12}$C($1^-_1$) calculated with the AMD+VAP \cite{KanadaEn'yo:2006ze}.
The orientation of an intrinsic state is chosen so as to satisfy 
$\langle x^2 \rangle \le \langle y^2 \rangle \le \langle z^2 \rangle$ and 
$\langle xy \rangle=\langle yz \rangle = \langle zx \rangle=0$. 
The horizontal and vertical axes are set to the $z$ and $y$ axes, respectively.
Densities are integrated with respect to the $x$ axis.
}
\end{figure}

In the previous study of $^{12}$C, totally, 23 AMD configurations
$\Phi^{\rm AMD}_{^{12}{\rm C}}({\bf Z}^{(k)})$ ($k=1,\dots, 23$) 
are obtained by the energy variation for 
$J(k)^{\pi(k)}_{n(k)}=0^+_1,0^+_2,0^+_3,1^+_1,2^+_1,2^+_2,2^+_3,
\cdots,1^-_1,2^-_1,3^-_1,\cdots$, and they are adopted 
as basis wave functions of the final wave functions of $^{12}$C. 
In the present $^{12}$C(AMD)+$\alpha$GCM calculation, we adopt 
only three 
basis wave functions to save the computational cost.
In order to take into account the ground and second $0^+$ states of $^{12}$C, 
we choose two basis wave functions of $J(k)^{\pi(k)}_{n(k)}=0^+_1,0^+_2$ for $k=1,2$. We also adopt 
the basis wave function of $J(k)^{\pi(k)}_{n(k)}=1^-_1$ for $k=3$. 
The intrinsic density of these three basis wave functions are shown 
in Fig.~\ref{fig:c12dense}.
The ground state has the compact structure of $3\alpha$ with a mixing of 
the $p_{3/2}$-shell closure component, while the $0^+_2$ and $1^-_1$ states show
developed $3\alpha$ cluster structures. 
 
\begin{figure}[th]
\epsfxsize=0.48\textwidth
\centerline{\epsffile{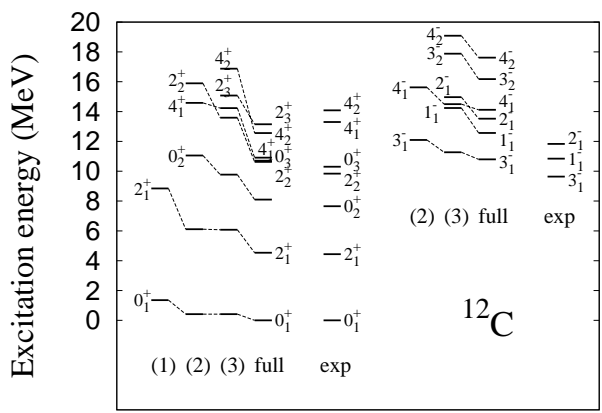}}
\caption{\label{fig:c12spe}
Energy spectra of $^{12}$C calculated with the AMD+VAP using 
three basis AMD wave functions $\Phi^{\rm AMD}_{^{12}{\rm C}}({\bf Z}^{(k)})$ 
obtained by the energy variation for $J(k)^{\pi(k)}_{n(k)}=0^+_1,0^+_2$ and $1^-_1$ with $k=1,2$ and 3, respectively, and that using the full 23 basis AMD wave functions. 
The $0^+_1$ energy calculated with the 23 bases is adjusted to 0 and the relative energies
are plotted. 
The energies calculated using (1) one basis ($k=1$), (2) two bases ($k=1,2$), (3) three bases ($k=1,2,3$),
and (full) the full bases are shown. 
The excitation energies of the experimental data are also shown.
}
\end{figure}

The energy levels of $^{12}$C obtained with the truncated model space of three bases are shown 
in Fig.~\ref{fig:c12spe} compared with those with 
full 23 basis wave functions and experimental ones. 
With the truncation, we get reasonable reproduction of the energy levels 
of many positive and negative parity states 
though the full 23 basis wave functions gives better results, in particular, for excited states.
The reason for $\sim$ 2 MeV higher energies of 
the $0^+_2$ and $1^-_1$ states with the 
three bases than those with the full bases is that these states gain their energy 
by the superposition of various  configurations of the $3\alpha$ cluster.

We also calculate the overlap ${\cal N}(^{16}{\rm O}(J^\pi_n);^{12}{\rm C}(0^+_n)+\alpha; d)$ 
of the $^{16}$O wave function obtained by the $^{12}$C(AMD)+$\alpha$GCM and the
$^{12}$C($0^+_n$)+$\alpha$ wave function having a certain distance $d$,
\begin{eqnarray}
&& \Phi_{^{12}{\rm C}(0^+_n)+\alpha}^{J\pi}(d)\equiv n_0
P^{J\pi}_{00}{\cal A}\{ \nonumber\\
&&\sum_{k}c^{0^+_n}_{^{12}{\rm C}}(K=0,k) P'^{0}_{00}
\Phi_{^{12}{\rm C}}^{\rm AMD}(-{\bf S}/4;{\bf Z}^{(k)})\nonumber\\
&&\times  \Phi_\alpha(3{\bf S}/4)\},\\
&&
{\cal N}(^{16}{\rm O}(J^\pi_n);^{12}{\rm C}(0^+_n)+\alpha; d)\nonumber\\
&&\label{eq:overlap} \equiv 
|\langle \Psi_{{\rm AMD}+\alpha{\rm GCM}}^{J^\pi_n} 
| \Phi_{^{12}{\rm C}(0^+_n)+\alpha}^{J\pi}(d)\rangle |^2.
\end{eqnarray}
Here $n_0$ is the normalization factor to satisfy 
\begin{equation}
|\langle \Phi_{^{12}{\rm C}(0^+_n)+\alpha}^{J\pi}(d)|
\Phi_{^{12}{\rm C}(0^+_n)+\alpha}^{J\pi}(d)\rangle|^2=1,
\end{equation}
and the $\Omega'$ integration in the operator $P'^{0}_{00}$ of the $J=0$ angular-momentum
projection of the subsystem is approximated 
by the sum of the finite number mesh points, $P'^{0}_{00}=\sum_{j}R^{\rm sub}(\Omega'_j)$. 
In the present work, we calculate the overlap only with the $^{12}{\rm C}(0^+_n)$-cluster wave function
because of the approximation with the finite points of the Euler angle $\Omega'_j$ for the 
rotation of subsystem $^{12}$C.

\subsection{Parameters in numerical calculations}
The width parameter $\nu$ of the $^{12}$C cluster 
is $\nu=0.19$ fm$^{-2}$ which was used in the previous work on $^{12}$C in 
Ref.~\cite{KanadaEn'yo:2006ze}. The width parameter of the $\alpha$ cluster is taken to be 
the same value  $\nu=0.19$ fm$^{-2}$ because 
the center of mass motion can be exactly extracted when a common
width parameter is used for all clusters.

For the inter-cluster distance between $^{12}$C and $\alpha$, 
six points $d_i$=$1.2,2.4,3.6\cdots,7.2$ fm are chosen.
The choice of $d_i\le 7.2$ fm corresponds to a kind of bound state approximation. 
In the angular-momentum projection of the total system, 
the integration of the Euler angle $\Omega=(\theta_1,\theta_2,\theta_3)$ 
is numerically performed by the 
summation of mesh points $(23,46,23)$ of the angles $(\theta_1,\theta_2,\theta_3)$.

For the intrinsic states of $^{12}$C labeled by $(k)$, three AMD configurations
are adopted.
For each intrinsic state $(k)$ at the distance $d_i$, seventeen rotated states $R^{\rm sub}(\Omega'_j)\Phi_{^{12}{\rm C}}^{\rm AMD}(-{\bf S}/4;{\bf Z}^{(k)})$ 
($j=1,\cdots,17$) are constructed.
The Euler angle $\Omega'=(\theta'_1,\theta'_2,\theta'_3)$ are chosen to be 
$\theta'_1=(0,\pi/4,\pi/2,3\pi/4,\pi)$ and $\theta'_2=(0,\pi/4,\pi/2,3\pi/4,\pi)$.
We omit the points $\theta'_2=(5\pi/4,3\pi/2,7\pi/4)$ in the region $\pi< \theta'_2< 2\pi$
to save the numerical cost. This is valid 
when the intrinsic state has the symmetry such as an isosceles triangle $3\alpha$ configuration. 
$\theta'_3$ is fixed to be 
$\theta'_3=0$ because the rotation $\theta'_3$ is effectively done 
by the $K$ projection in the angular-momentum projection of the total system
because of the rotational invariance of the $\alpha$ cluster.
As for the $K$-mixing, we truncate the $|K| \ge 4 $ components.

\section{Results}\label{sec:results}

\subsection{Effective nuclear interaction}
In the present calculation of the $^{12}$C(AMD)+$\alpha$GCM, we use
the same effective nuclear interaction with the same parameters 
as those used
in the previous calculation of $^{12}$C\cite{KanadaEn'yo:2006ze}. 
It is the MV1 force \cite{MVOLKOV} 
 for the central force 
supplemented by the two-body spin-orbit force with the two-range Gaussian form 
same as that in the G3RS force \cite{LS}.
The Coulomb force is approximated using a seven-range
Gaussian form. 
The Majorana, Bartlett, 
and Heisenberg parameters in the MV1 force 
are $m=0.62$, $b=0$, and $h=0$, respectively, and the 
spin-orbit strengths are taken to be $u_{I}=-u_{II}=3000$ MeV.

\subsection{Energy levels of $0^+$ states}

In the preceding studies\cite{Funaki:2008gb,Funaki:2010px,suzuki76,Libert80,supp80,Fukatsu92}, 
developed cluster structures were suggested 
in excited $0^+$ states of $^{16}$O. It is considered that 
the ground state of $0^+_1$ is dominated by the doubly closed-shell structure,
while the $0^+_2$ state has the $^{12}$C($0^+_1$)+$\alpha$ structure. 
The $0^+_3$ state is suggested to have the $^{12}$C($2^+_1$)+$\alpha$ component.
In the $4\alpha$-OCM calculation, it was suggested that the $0^+_4$ 
mainly has the $^{12}$C($0^+_1$)+$\alpha$  structure with higher nodal behavior of 
$\alpha$ cluster around $^{12}$C and the $0^+_5$ contains the  
$^{12}$C($1^-_1$)+$\alpha$ component.
In the study with the $4\alpha$-OCM calculation by Funaki {\it et al.}\cite{Funaki:2008gb,Funaki:2010px},
the $0^+_6$ state having the 
$^{12}$C($0^+_2$)+$\alpha$ structure was suggested and regarded as the $4\alpha$ cluster gas state.
They proposed that the experimental $0^+_6$ state at 15.1 MeV is a candidate 
for the $4\alpha$ cluster gas state.

\begin{figure}[th]
\epsfxsize=0.48\textwidth
\centerline{\epsffile{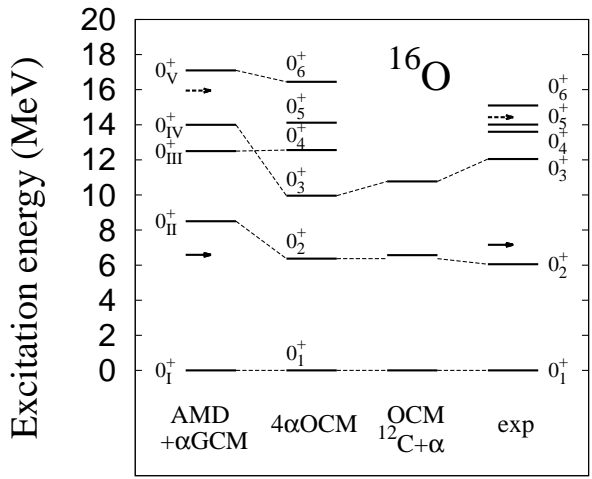}}
\caption{\label{fig:o16spe}
Excitation energies of $0^+$ states in $^{16}$O
calculated with the present $^{12}$C(AMD)+$\alpha$GCM (AMD+$\alpha$GCM) 
and those of the 4$\alpha$-OCM and 
$^{12}$C+$\alpha$-OCM from Refs.~\cite{Funaki:2008gb,suzuki76}.
The experimental energy levels of 
$0^+$ states are taken from Refs.~\cite{Tilley:1993zz,Wakasa:2007zza}.
The $^{12}$C($0^+_1$)+$\alpha$ and $^{12}$C($0^+_2$)+$\alpha$ threshold energies are plotted 
by solid and dashed arrows, respectively.}
\end{figure}

The energy levels of $0^+$ states of $^{16}$O up to the fifth $0^+$ state 
calculated with the 
$^{12}$C(AMD)+$\alpha$GCM calculation are shown in Fig.~\ref{fig:o16spe} compared with the 
experimental data. The theoretical energy levels with other theoretical calculations,
$4\alpha$-OCM\cite{Funaki:2008gb,Funaki:2010px} and $^{12}$C+$\alpha$-OCM\cite{suzuki76}, are also shown.

In the present result, 
the ground state ($0^+_I$) 
has mainly the doubly closed-shell structure with less development of cluster, while 
the second $0^+$ state ($0^+_{II}$) is described mainly 
by the developed $^{12}$C($0^+_1$)+$\alpha$ structure. 
The cluster structure of the $0^+_{II}$ state is
consistent with that of the preceding works and can be assigned to the experimental 
$0^+_2$ state at 6.05 MeV. 
Above the second $0^+$ state, we obtain 
the third $0^+$ state ($0^+_{III}$) having further developed $^{12}$C($0^+_1$)+$\alpha$  structure and the forth $0^+$ state ($0^+_{IV}$) showing a feature of 
the $^{12}$C($2^+_1$)+$\alpha$ structure. The features of the $0^+_{III}$ and $0^+_{IV}$ states are 
consistent with the $0^+_4$ and $0^+_3$ states in the $4\alpha$-OCM calculation \cite{Funaki:2008gb,Funaki:2010px,Yamada:2011ri},
respectively.  
The ordering of the $0^+_{III}$ and $0^+_{IV}$
is opposite to that of the 4$\alpha$-OCM calculation. If we assign the $0^+_{IV}$ state to the
$0^+_3$ at 12.05 MeV, the experimental 
level spacing between the $0^+_2$ and $0^+_3$ state is reproduced well by the present 
calculation. 
For the $0^+_{III}$ state, the dominant $^{12}$C($0^+_1$)+$\alpha$ structure with an $\alpha$ far from
the $^{12}$C($0^+_1$) core is consistent with the $0^+_4$ state in the 4$\alpha$-OCM calculation which 
is assigned to the $0^+_4$ state at 13.6 MeV from its relatively large width. 
The present calculation is a bound state approximation, and therefore it is difficult to 
discuss the width. Moreover, stability of this state should be checked carefully by taking into account 
mixing of continuum states. 

In the present calculation, we obtain the fifth $0^+$ state ($0^+_{V}$) 
having the developed $^{12}$C($0^+_2$)+$\alpha$ structure. As discussed later, 
it has a large $^{12}$C($0^+_2$)+$\alpha$ component 
with an $\alpha$ cluster 
moving around the $^{12}$C($0^+_2$) cluster in the $S$-wave channel and it is consistent with
the structure of the 4$\alpha$ cluster gas state in the $0^+_6$ suggested by the $4\alpha$-OCM calculation. 
The $0^+_{V}$ state has a significant $\alpha$-cluster amplitude about 
$4\sim 5$ fm far from 
the $^{12}$C$(0^+_2)$ core. This is the same region of the $\alpha$-cluster amplitude in the 
$^{12}$C$(0^+_2)$. Considering that the $^{12}$C$(0^+_2)$ has the $3\alpha$ cluster gas feature 
and the fourth $\alpha$ is moving in $S$-wave in the same region of 3 $\alpha$ clusters, 
the $0^+_{V}$ state can be regarded as a 4$\alpha$ cluster gas state similar to 
the $3\alpha$ cluster gas in the $^{12}$C$(0^+_2)$.

\begin{table}[ht]
\caption{
\label{tab:radii}
The charge radii($R_c$), $E0$ transition matrix elements($M(E0)$), and the 
ratio($P_{\rm E.W.}$) to
the EWSR of the isoscalar $E0$ transition that is 1/4 of the isoscalar monopole
EWSR. 
The theoretical values are those calculated with the present $^{12}$C(AMD)+$\alpha$GCM, 
the $4\alpha$-OCM\cite{Yamada:2011ri}, the $^{12}$C+$\alpha$-OCM\cite{suzuki76}.
The charge radius of a proton 0.887 fm \cite{atomic-data} is used to evaluate 
the charge radii of $0^+$ states in $^{16}$O. 
Experimental data are taken from 
Refs.~\cite{Tilley:1993zz,atomic-data,Ajzenberg}.}
\begin{center}
\begin{tabular}{ccccc}
\hline
\hline
 state & $E_x$ (MeV) & $R_c$ (fm) & $M(E0)$ & $P_{\rm E.W.}$  ($\%$) \\
 \multicolumn{5}{c}{ $^{12}$C(AMD)+$\alpha$GCM } \\
$0^+_I$	&	0.0 	&	2.9 	&		&		\\
$0^+_{II}$	&	8.5 	&	3.5 	&	4.0 	&	5.4$\%$	\\
$0^+_{III}$	&	12.5 	&	3.9 	&	3.5 	&	6.4$\%$	\\
$0^+_{IV}$	&	14.0 	&	3.5 	&	6.0 	&	20$\%$	\\
$0^+_{V}$	&	17.1 	&	3.8 	&	1.4 	&	1.4$\%$	\\
\hline
\multicolumn{5}{c}{exp. } \\ 									
$0^+_1$	&	0	&	2.70	&		&		\\
$0^+_2$	&	6.05	&		&	3.55(0.21)	&	3.5$\%$	\\
$0^+_3$	&	12.05	&		&	4.03(0.09)	&	9.1$\%$	\\
$0^+_4$	&	13.6	&		&		&		\\
$0^+_5$	&	14.01	&		&	3.3(0.7)	&	6.3$\%$	\\
$0^+_6$	&	15.1	&		&		&		\\
\multicolumn{5}{c}{$4\alpha$-OCM} \\ 							
$0^+_1$	&	0	&	2.7	&		&		\\
$0^+_2$	&	6.37	&	3	&	3.9 	&	4$\%$	\\
$0^+_3$	&	9.96	&	3.1	&	3.9 	&	6.3$\%$	\\
$0^+_4$	&	12.56	&	4	&	2.4 	&	3$\%$	\\
$0^+_5$	&	14.12	&	3.1	&	2.6 	&	3.9$\%$	\\
$0^+_6$	&	16.45	&	5.6	&	1.0 	&	0.7$\%$	\\
\multicolumn{5}{c}{$^{12}$C+$\alpha$-OCM} \\
$0^+_1$	&	0	&	2.5	&		&		\\
$0^+_2$	&	6.57	&	2.9	&	3.88	&	4.8$\%$	\\
$0^+_3$	&	10.77	&	2.8	&	3.5	&	6.4$\%$	\\
\hline
\hline
\end{tabular}
\end{center}
\end{table}
The root-mean-square charge radii and monopole transition matrices $M(E0)$ for 
the $0^+$ states are shown in Table \ref{tab:radii}. 
The excited states tend to have large r.m.s. charge radii due to developed cluster structures 
compared with that of the ground state.
In particular, the $0^+_{III}$ state with the higher nodal $^{12}$C$(0^+_1)$+$\alpha$ structure and
$0^+_{V}$ state with the $^{12}$C$(0^+_2)$+$\alpha$ structure have about 1 fm larger radii 
than the ground state. The radii of the 
$0^+_{V}$ state is smaller than the $0^+_6$ state of the $4\alpha$-OCM calculation.
It may come from the smaller radius of $^{12}$C($0^+_2$) with the 3-basis AMD+VAP calculation 
than that with the $3\alpha$-OCM calculation\cite{Yamada:2005ww}.
Namely, the r.m.s. matter radius of  $^{12}$C($0^+_2$) is 3.2 fm in the 3-basis AMD+VAP result
(3.3 fm in the full 23-basis AMD+VAP) and
4.31 fm in the $3\alpha$-OCM calculation. 

Those excited states with  developed cluster structures also have significant 
monopole transition strength from the ground state. The transition strength to the $0^+_V$ state
is relatively smaller than those to the lower $0^+$ states.  
The present result is consistent with that of the $4\alpha$-OCM calculation 
in Ref.~\cite{Yamada:2011ri}.
Detailed discussion of isoscalar monopole excitations is given in the next section.

\subsection{$E2$ transition strength and band assignment}
\begin{figure}[th]
\epsfxsize=0.4\textwidth
\centerline{\epsffile{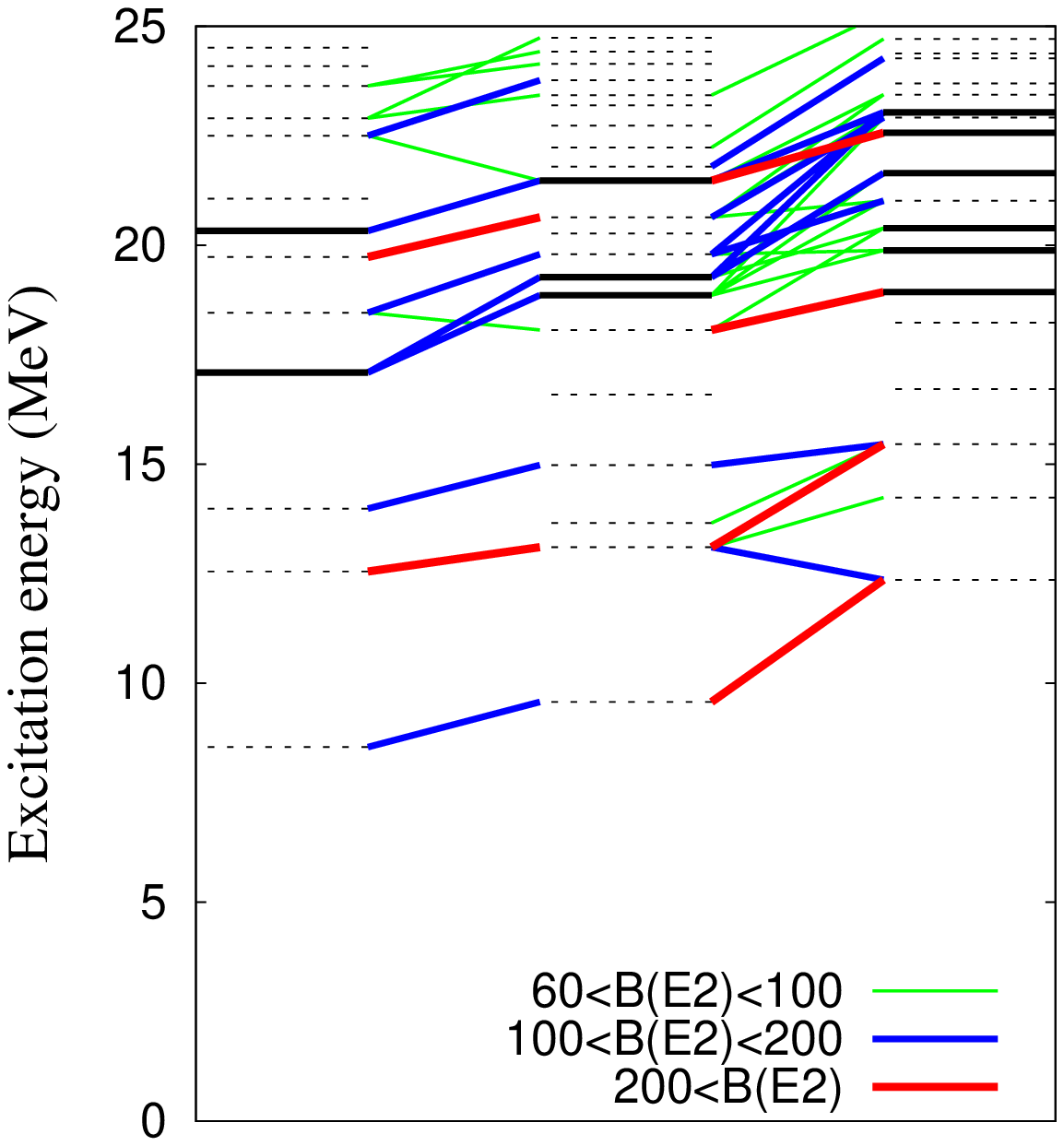}}
\epsfxsize=0.4\textwidth
\centerline{\epsffile{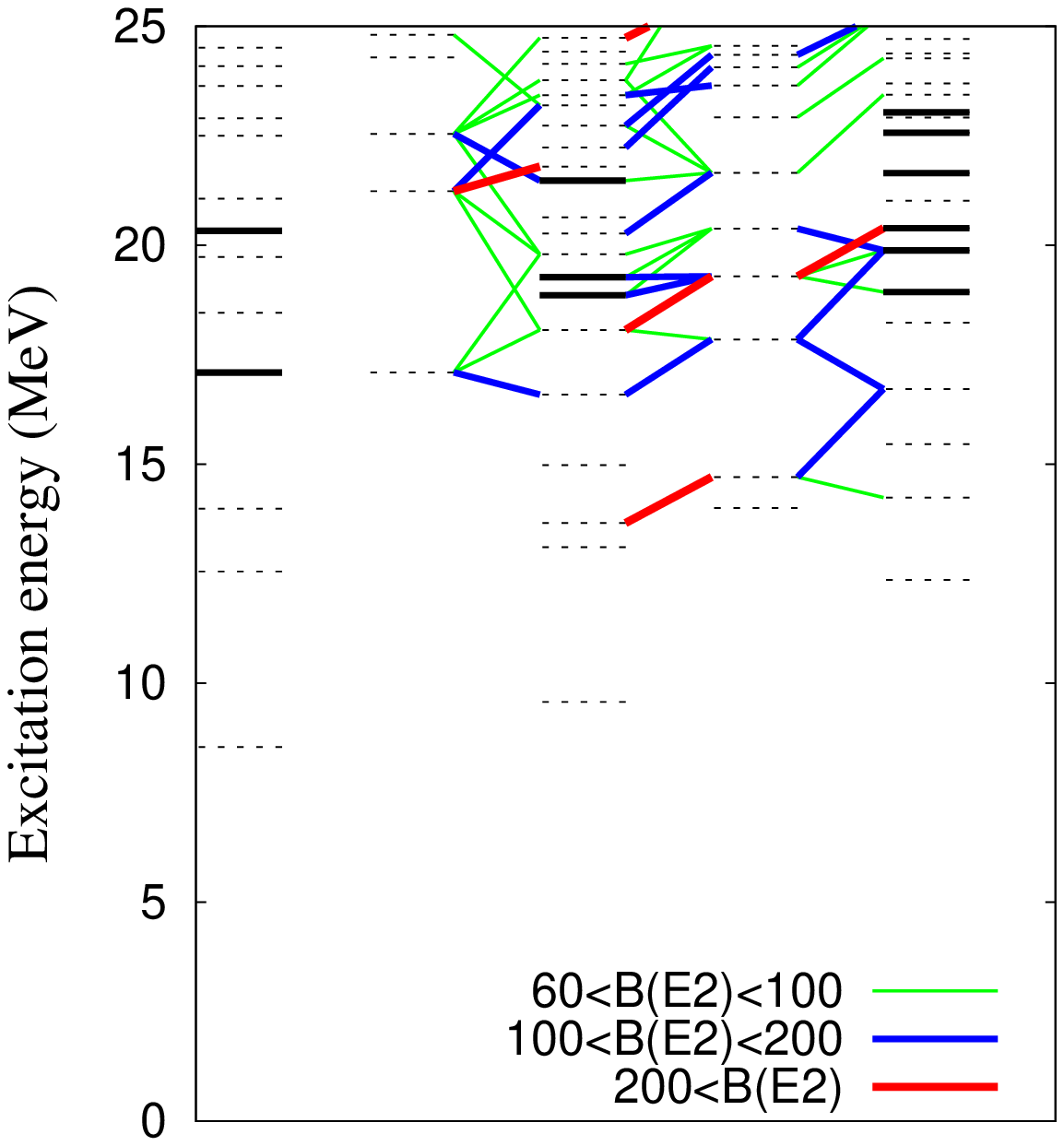}}
\caption{\label{fig:be2}
(Color online)
Calculated $E2$ transition strength in $^{16}$O
obtained with the $^{12}$C(AMD)+$\alpha$GCM.
States having strong $E2$ transition with $60<B(E2)<100$ e$^2$fm$^4$, 
$100<B(E2)<200$ e$^2$fm$^4$, and $200<B(E2)$ e$^2$fm$^4$ are connected 
by green, blue, and red lines, respectively.
(Upper) $J^\pi=0^+,2^+,4^+$ spectra (dashed) and
 $4^+\rightarrow 2^+$ and $2^+\rightarrow 0^+$ transitions(red, blue, and green lines).
(Lower)$J^\pi=0^+,1^+,2^+,3^+,4^+$ spectra and $J^+\rightarrow (J-1)^+$ transitions.
The solid lines are energy levels having significant $^{12}$C($0^+_2$)+$\alpha$
component.}
\end{figure}

As mentioned above, the present result suggests the $^{12}$C($0^+_2$)+$\alpha$ structure
in the $0^+_V$ state which is regarded as the candidate for the 4$\alpha$ cluster gas state.
By analyzing the calculated $E2$ transition strength, we consider rotational band members 
from the  $^{12}$C($0^+_2$)+$\alpha$ structure.
The calculated $E2$ transition strength is shown in Fig.\ref{fig:be2}.
The experimental and theoretical $B(E2)$ values for low-energy states 
are listed in Table \ref{tab:be2}. 
For the lowest $^{12}$C($0^+_1$)+$\alpha$ cluster band  consisting of 
the $0^+_2$, $2^+_1$, and $4^+_1$ states, 
the present $^{12}$C(AMD)+$\alpha$GCM calculation reproduce reasonably the strong intra-band $E2$ 
transitions within a factor two.  Twice larger $B(E2)$ values than the experimental data 
may suggest $\sim 20\%$ overestimation of the r.m.s. radii of these states which may come from
the higher energy position relative to the $^{12}$C($0^+_1$)+$\alpha$ threshold.
For the  second $2^+$ state($2^+_{II}$), which can be understood as 
the rotational member of the $0^+_{III}$, 
the strong $E2$ transition to the ground state is inconsistent with the 
experimental data. This results suggest again that the $0^+_{III}$ and $2^+_{II}$ states should be assigned 
to higher $0^+$ and $2^+$ states instead of the $0^+_3$ and $2^+_2$ states.
If we assign the third and the fourth $2^+$ states ($2^+_{III}$ and $2^+{IV}$) obtained in the
present calculation to the $2^+_2$ and $2^+_3$ states, the calculated $B(E2)$ values are in reasonable agreement with the experimental ones.

\begin{table}[ht]
\caption{
\label{tab:be2}
$E2$ transition strength in $^{16}$O. 
$B(E2)$ values  calculated with the present $^{12}$C(AMD)+$\alpha$GCM
and those with the $^{12}$C+$\alpha$-OCM\cite{suzuki76}.
Experimental data are taken from  Ref.~\cite{Tilley:1993zz}.}
\begin{center}
\begin{tabular}{cccc}
\hline
\hline
initial & final & \multicolumn{2}{c} {$B(E2)$ (e$^2$fm$^4$)} \\
 & & exp. & Ref.~\cite{suzuki76} \\
$2^+$(6.92)	&	$0^+(0)$	&	7.4$\pm$0.2	&	2.48	\\
$2^+$(6.92)	&	$0^+$(6.05)	&	65$\pm$7	&	60.1	\\
$2^+$(9.84)	&	$0^+(0)$	&	0.07$\pm$0.007	&	0.489	\\
$2^+$(9.84)	&	$0^+$(6.05)	&	2.9$\pm$0.7	&	4.64	\\
$2^+$(11.5)	&	$0^+(0)$	&	3.6$\pm$1.2	&	1.43	\\
$2^+$(11.5)	&	$0^+$(6.05)	&	7.4$\pm$1.2	&	1.38	\\
$4^+$(10.4)	&	$2^+$(6.92)	&	156$\pm$14	&	96.2	\\
\hline
 & & present &  \\						
$2^+_{I}$	&	$0^+_{I}$	&	3.2	&		\\
$2^+_{I}$	&	$0^+_{II}$	&	177	&		\\
$2^+_{II}$	&	$0^+_{I}$	&	45	&		\\
$2^+_{II}$	&	$0^+_{II}$	&	2.3	&		\\
$2^+_{III}$	&	$0^+_{I}$	&	0.08	&		\\
$2^+_{III}$	&	$0^+_{II}$	&	1.4	&		\\
$2^+_{IV}$	&	$0^+_{I}$	&	3.1	&		\\
$2^+_{IV}$	&	$0^+_{II}$	&	0.1	&		\\
$4^+_{I}$	&	$2^+_{I}$	&	290	&		\\
\hline
\end{tabular}
\end{center}
\end{table}

In the energy region around $E_x\sim 20$ MeV, we find $2^+$ states and $4^+$ states
having rather strong (sequential) $E2$ transition strength toward the $0_V$ state.
In this energy region, there are several $2^+$ and $4^+$ states having 
non-negligible component of the $^{12}$C($0^+_2)$+$\alpha$ structure. 
 We also obtain another $0^+$ state
with some $^{12}$C($0^+_2)$+$\alpha$ component at 20.3 MeV, 
a few MeV above the $0_V$ state.
In Fig.~\ref{fig:be2}, the energy levels of these states are shown by solid lines.
$E2$ transition strength is fragmented among them as shown in the figure. 

\begin{figure}[th]
\epsfxsize=0.3\textwidth
\centerline{\epsffile{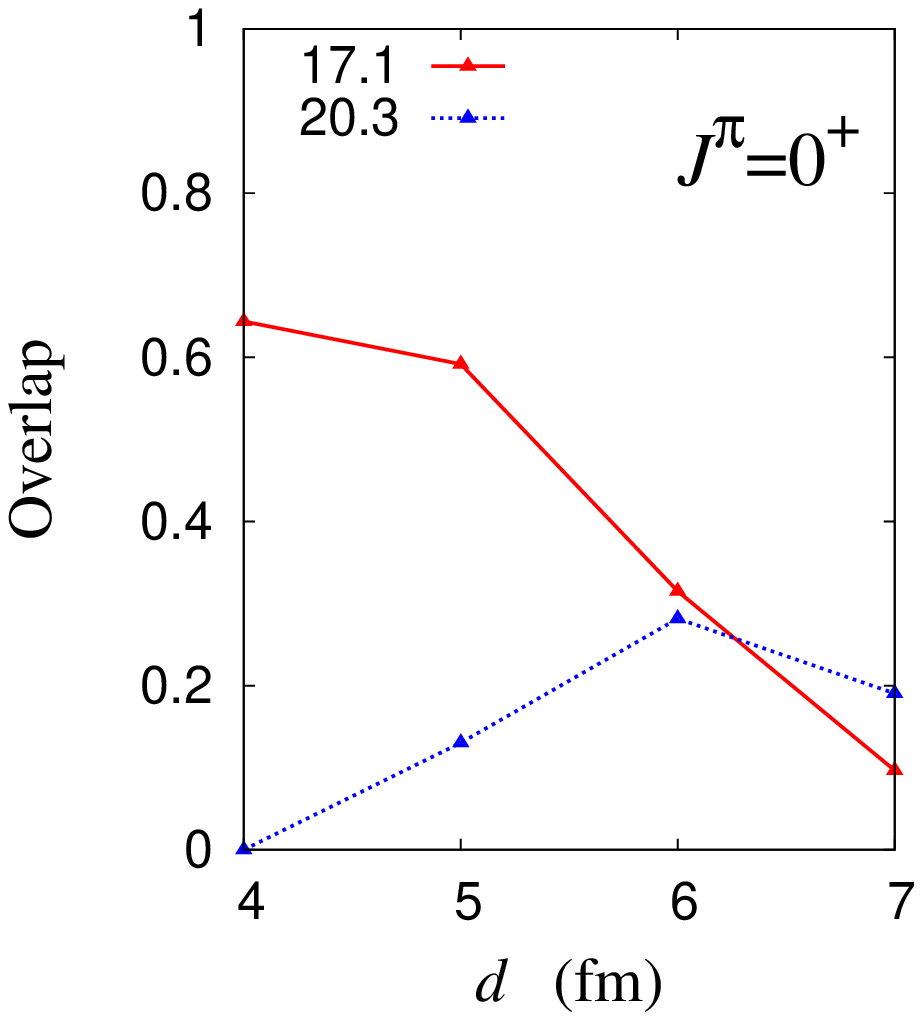}}
\epsfxsize=0.3\textwidth
\centerline{\epsffile{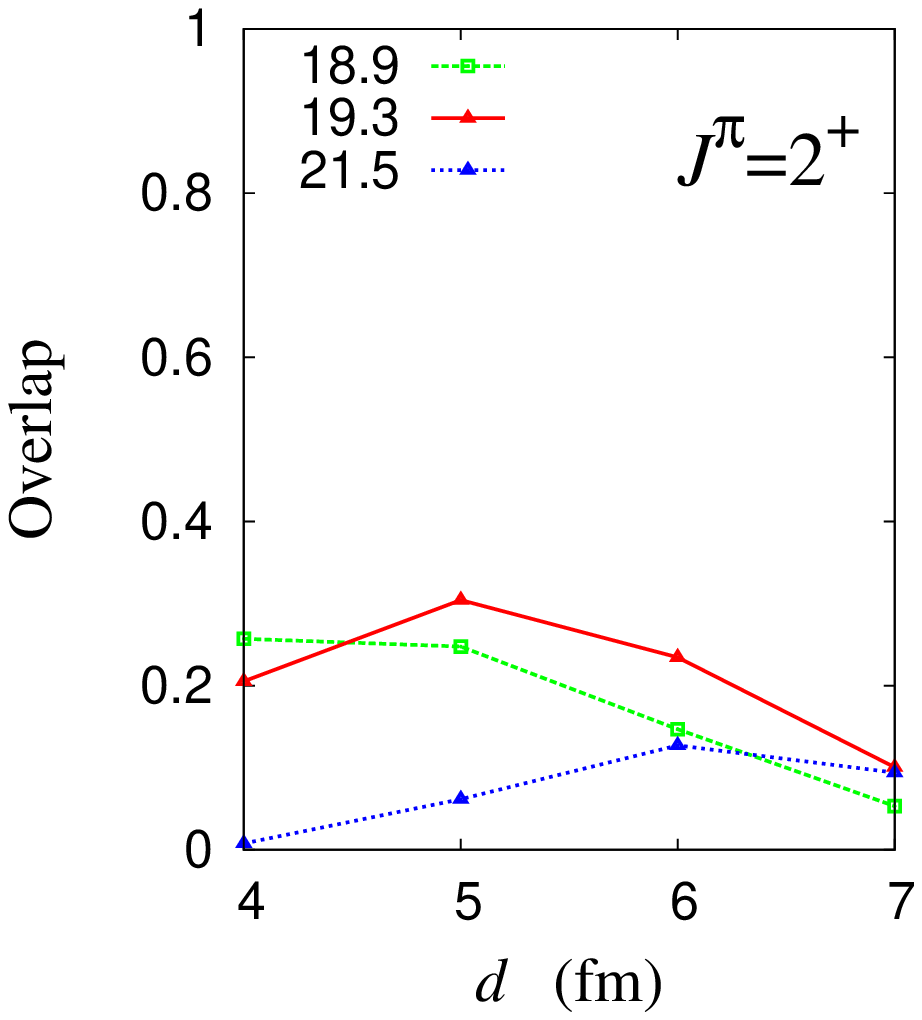}}
\epsfxsize=0.3\textwidth
\centerline{\epsffile{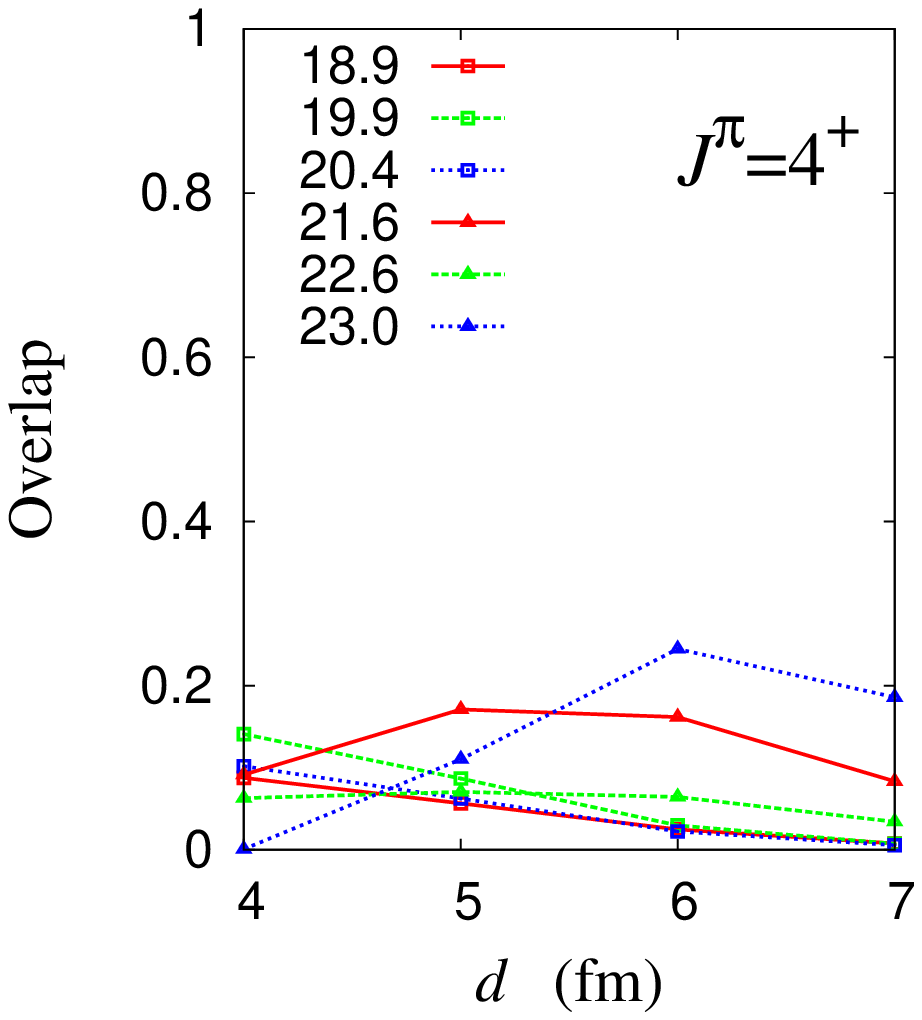}}
\caption{
\label{fig:overlap2}
(Color online) The overlap of the excited 
$^{16}$O states
with the $^{12}$C($0^+_2)$+$\alpha$ wave function 
as a function the inter-cluster distance $d$ 
defined in Eq.~\ref{eq:overlap}.
The calculated overlap for the 
$0^+$ states at 17.1 and 20.3 MeV, 
$2^+$ states at 18.9, 19.3, and 21.5 MeV, and
$4^+$ states at 18.9, 19.9, 20.4, 21.6, 22.6 and 23.0 MeV, which have 
significant $^{12}$C($0^+_2)$+$\alpha$
component, is shown.}
\end{figure}

Figure \ref{fig:overlap2} shows the overlap of those states with 
the $^{12}$C($0^+_2)$+$\alpha$ wave function as function of the inter-cluster distance $d$.
The $0^+_V$ state has more than 60$\%$
$^{12}$C($0^+_2)$+$\alpha$ component at $d_\alpha=4-5$ fm. As the spin increases, 
the $^{12}$C($0^+_2)$+$\alpha$ component decreases and seems scattered into several 
$2^+$ and $4^+$ states. It may imply that the structure change, in other words, the 
state mixing occurs in the rotation of the $^{12}$C($0^+_2)$+$\alpha$ structure. 

\begin{figure}[th]
\epsfxsize=0.3\textwidth
\centerline{\epsffile{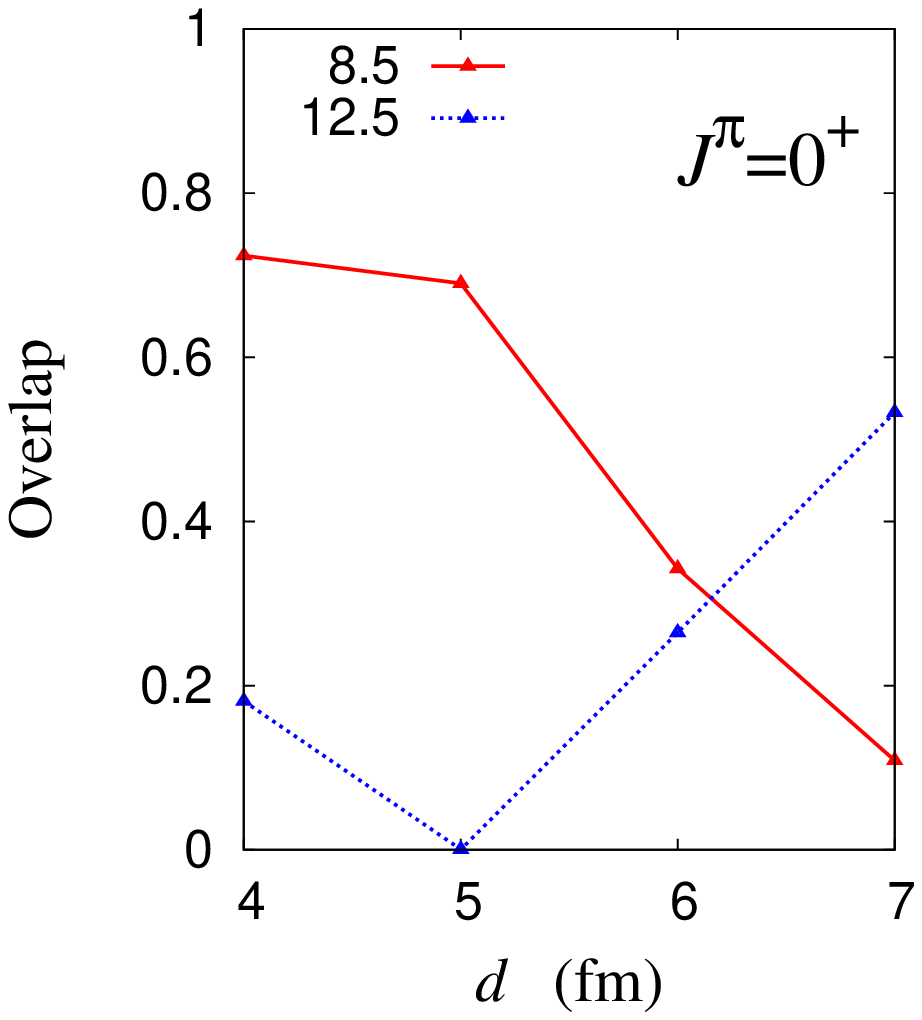}}
\epsfxsize=0.3\textwidth
\centerline{\epsffile{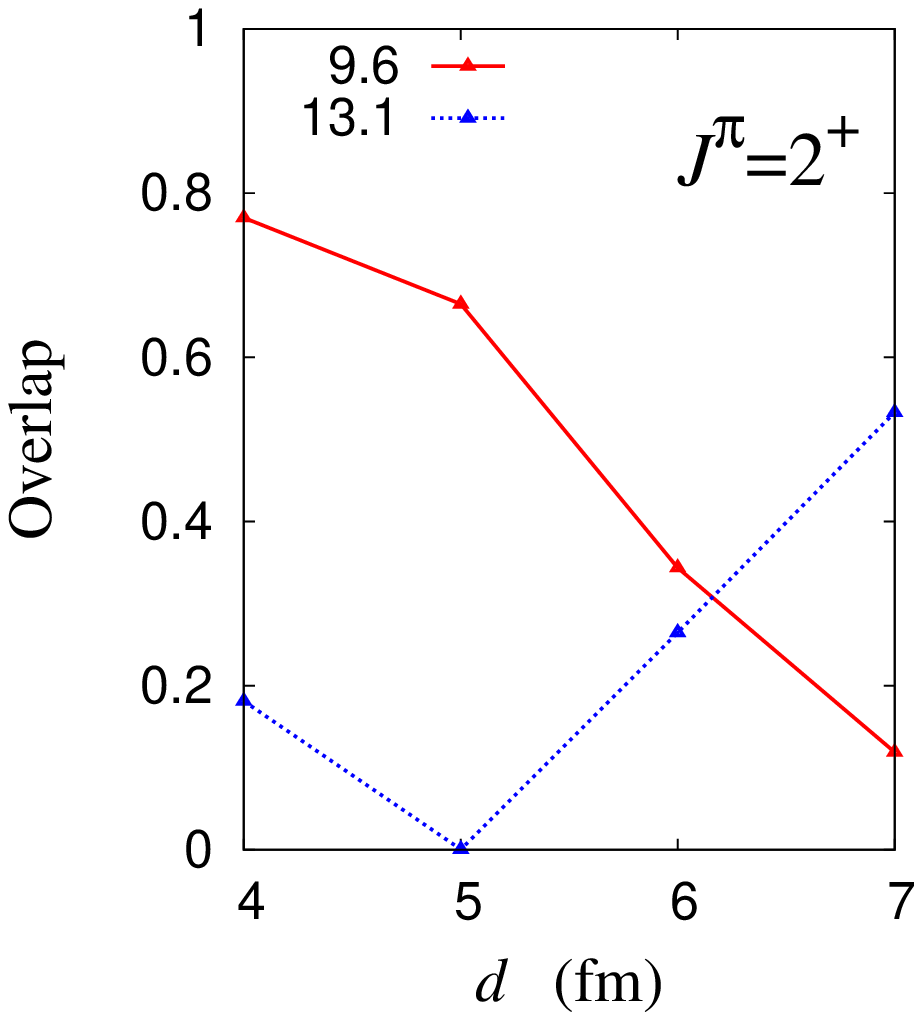}}
\epsfxsize=0.3\textwidth
\centerline{\epsffile{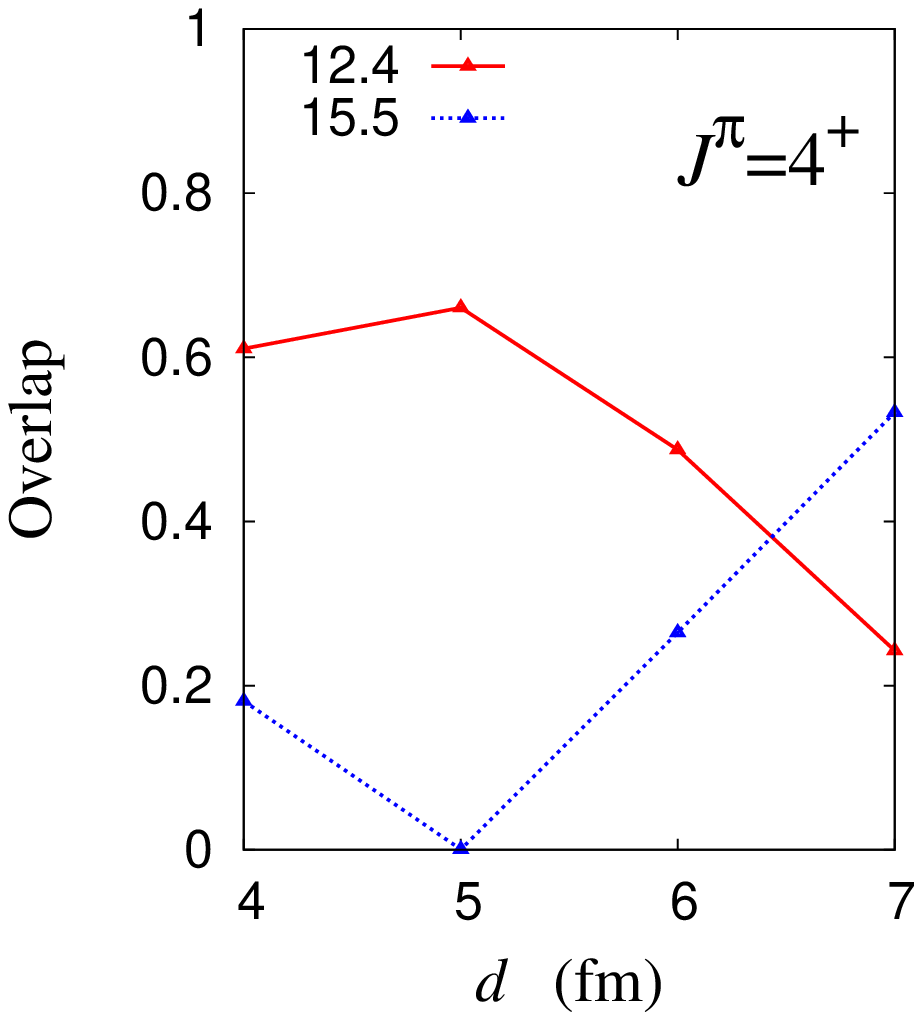}}
\caption{\label{fig:overlap1}
(Color online) The overlap of the $^{16}$O states 
with the $^{12}$C($0^+_1)$+$\alpha$ wave function as a function the inter-cluster distance $d$ 
defined in Eq.~\ref{eq:overlap}.
The overlap for the $0^+_{II}$ at 8.5 MeV, $2^+$ at 9.6 MeV, and $4^+$ at 12.4 MeV 
in the $^{12}$C($0^+_1)$+$\alpha$ band, and the $0^+_{III}$ at 12.5 MeV, $2^+$ at 13.1 MeV
and $4^+$ at 15.5 MeV in the higher nodal $^{12}$C($0^+_1)$+$\alpha$ band is shown.
}
\end{figure}

We also show in Fig.~\ref{fig:overlap1} the overlap with the  $^{12}$C($0^+_1)$+$\alpha$ wave function 
in the member states of the rotational bands starting from the $0^+_{II}$ and  $0^+_{III}$ states.
As seen in the figure, the lower band build on the $0^+_{II}$ has the 
$^{12}$C($0^+_1)$+$\alpha$ structure and higher band from the $0^+_{III}$ shows 
the higher nodal feature of the $^{12}$C($0^+_1)$+$\alpha$ structure.
The overlap with the $^{12}$C($0^+_1)$+$\alpha$ wave function in these states does not 
depend so much on the spin and it is still 
significant even in the $4^+$ states.

Thus, the situation is quite different between the $^{12}$C($0^+_1)$+$\alpha$ cluster bands 
and the $^{12}$C($0^+_2)$+$\alpha$ bands. 
The instability of the $^{12}$C($0^+_2)$+$\alpha$ states in the rotation is not surprising 
because the $^{12}$C($0^+_2$) cluster is considered to be the $3\alpha$ cluster gas and such a gas state 
should not be a rigid but fragile one differently from the $^{12}$C($0^+_1$) cluster.

\begin{figure}[th]
\epsfxsize=0.4\textwidth
\centerline{\epsffile{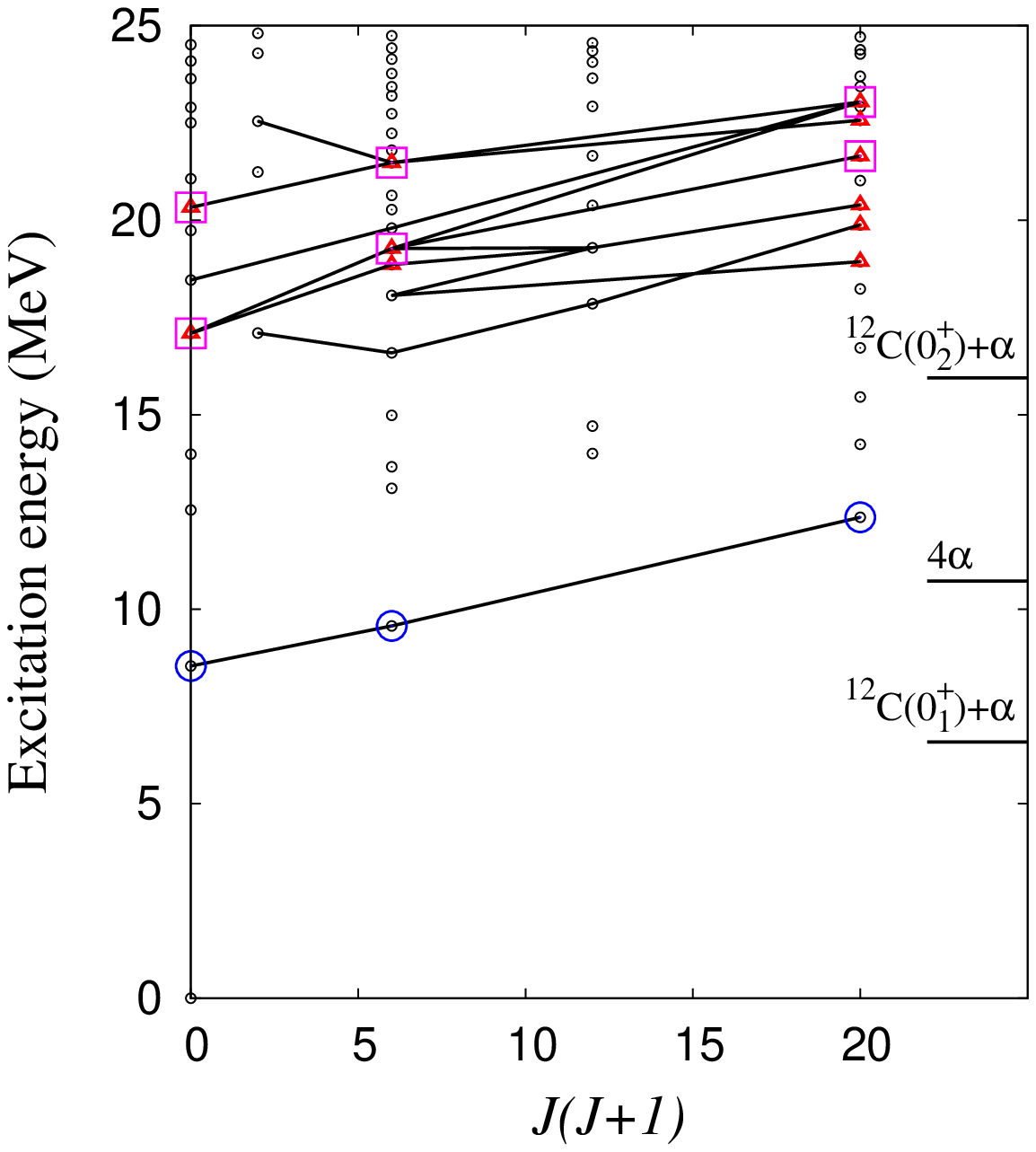}}
\epsfxsize=0.4\textwidth
\centerline{\epsffile{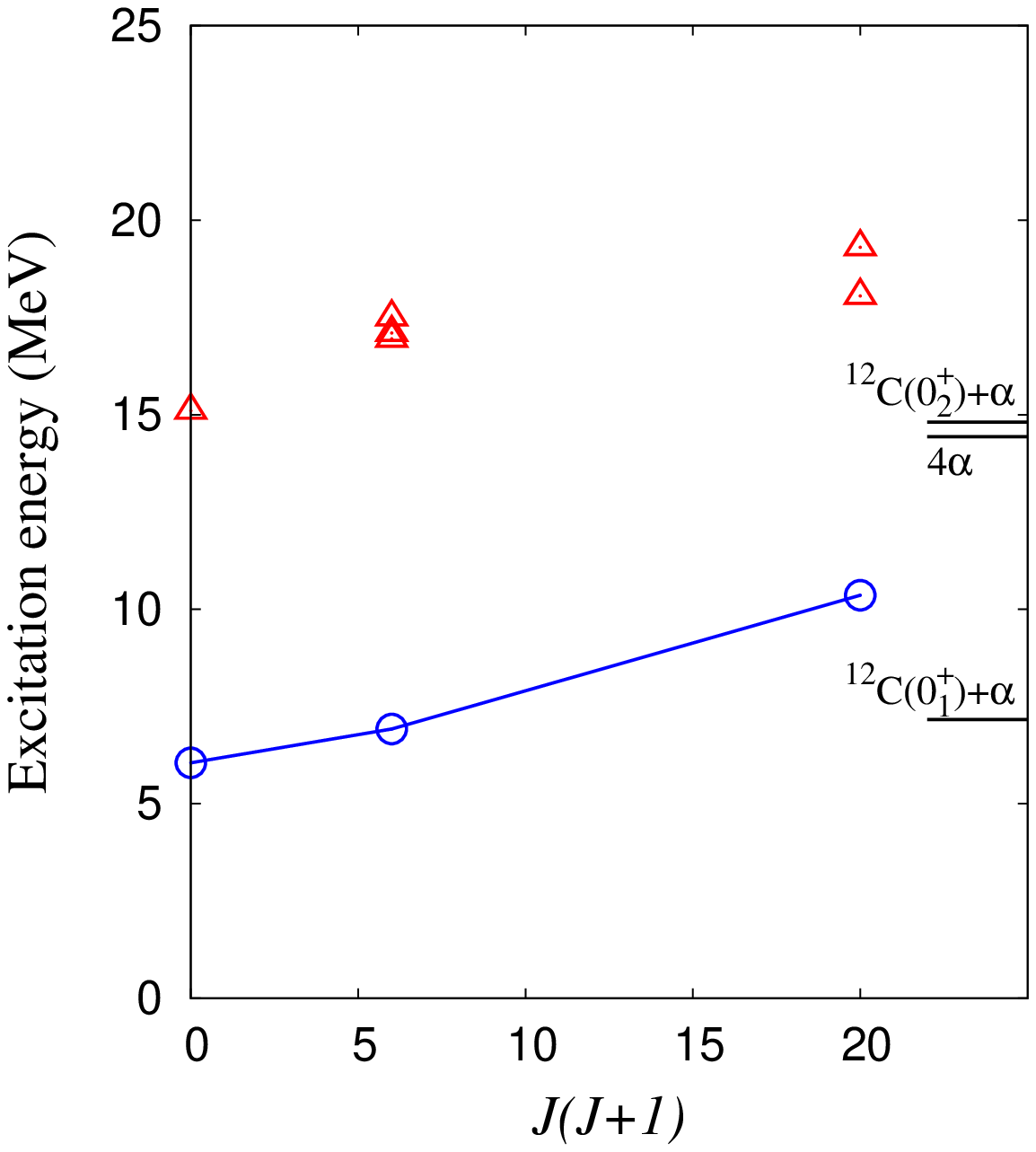}}
\caption{\label{fig:band}(Color online)
Upper: theoretical excitation levels of $^{16}$O 
calculated with the $^{12}$C(AMD)+$\alpha$GCM (circles). 
Excited states with significant $^{12}$C($0^+_2$)+$\alpha$ component
are shown by red triangles and squares. States connected by the lines are 
those which can be connected to the $4^+$ states with the 
$^{12}$C($0^+_2$)+$\alpha$ component by
the strong (sequential) transitions.
 Squares indicate possible band assignment
for the $^{12}$C($0^+_2$)+$\alpha$ cluster states.
Circles shows the band members of the
$^{12}$C($0^+_1$)+$\alpha$ cluster structure staring from the $0^+_{II}$ state. 
Lower: experimental excitation energies of the candidate states for the 
 $^{12}$C($0^+_2$)+$\alpha$ cluster states observed by the 
$^{12}$C($^{12}$C,$^8$Be+$^8$Be) and the $^{12}$C($^{16}$O, $4\alpha$)
reactions\cite{Chevallier:1967zz,Freer:1995zza}, and 
those of the band members of the 
$^{12}$C($0^+_1$)+$\alpha$ structure\cite{Tilley:1993zz}.}
\end{figure}

Consequently, it is difficult to clearly identify the band members of the $^{12}$C($0^+_2$)+$\alpha$ cluster state, however, 
considering the 
relatively strong $E2$ transition strength and similarity of the 
$d$-dependence of the $^{12}$C($0^+_2)$+$\alpha$ overlap,
we propose a possible assignment that the $2^+$ state at 19.3 MeV and $4^+$ at 21.6 MeV 
can be regarded as the band members from the $0^+_V$ state, and 
the $2^+$ state at 21.5 MeV and $4^+$ at 23.0 MeV are interpreted as members of the band  
staring from the $0^+$ state at 21.3 MeV.
The excitation energies are plotted as a function of the spin $J(J+1)$ in Fig.~\ref{fig:band}.
Square points indicates the assigned states, triangles shows 
the states with significant $^{12}$C($0^+_2)$+$\alpha$ component, and circles 
do the rotational members of the $^{12}$C($0^+_1$)+$\alpha$ band starting
from the $0^+_{II}$ state. Reflecting the structure change, the slope of the energy 
for $J(J+1)$ does not show the linear dependence but it becomes gentle 
with the increase of spin.
We also show in Fig.~\ref{fig:band} 
the experimental energy levels of the excited states observed 
in the $^{12}$C($^{12}$C,$^8$Be+$^8$Be) and the $^{12}$C($^{16}$O, $4\alpha$)
reactions\cite{Chevallier:1967zz,Freer:1995zza}, 
which are considered to be candidates for the $^{12}$C($0^+_2$)+$\alpha$ cluster 
states \cite{Ohkubo:2010zz}.  
The calculated energies of the $^{12}$C($0^+_2)$+$\alpha$ states 
measured from the $^{12}$C($0^+_2$)+$\alpha$ threshold in the present result 
qualitatively agree with
those of the experimental data. 

\section{Isoscalar monopole excitation}\label{sec:discussions}
As discussed recently, 
isoscalar monopole (ISM) excitation in the low-energy part gives important 
information on cluster structures of
excited states in light nuclei\cite{Yamada:2011ri,Yamada-monopole}.
As well known, the isoscalar giant monopole resonances (ISGMR) in heavy nuclei 
have been observed as 
a single peak and described by the collective breathing mode. The systematics of 
the peak position has been discussed in association with the nuclear compressibility.
In light nuclei such as $^{12}$C and $^{16}$O, however, it has been revealed 
by the $(e,e')$ and $(\alpha,\alpha')$ scattering experiments
\cite{Krison76,Lui:2001xh} that 
the ISM strength is strongly fragmented and significant fraction of the energy-weighted sum rule concentrates on a few states in a low-energy region.
Recently, Yamada {\it et al.} discussed the ISM excitation in $^{16}$O and showed that 
the significant ISM strength at the low-energy  part up to $E_x \sim$ 16 MeV can be described well 
by the monopole excitation to the cluster states\cite{Yamada:2011ri}. It was argued that 
two different types of monopole excitation exist in $^{16}$O, that is, 
the monopole excitation to cluster states dominating the strength
in the lower-energy part and that of the mean-field type $1p$-$1h$ excitation 
yielding the strength in the higher-energy part $16 \le E_x \le 40$ MeV. 

In principle, these two modes are not decoupled from but should couple
to each other because the cluster excitation partially involves the $1p$-$1h$ excitation. 
Indeed, the ISGMR peak position can be approximately described by the 
breathing mode of the radial motion of four $\alpha$ clusters \cite{Furuta:2010ad}.
Therefore, it is expected that the low-lying cluster states feed the strength of a part of the 
ISGMR strength originally concentrating at the higher energy region.

Although the cluster model calculations such as 
the $4\alpha$-OCM are useful to describe the cluster excitation, they are not enough 
to describe the mean-field type $1p$-$1h$ excitation because frozen 4 $\alpha$ clusters are assumed.
Also the present calculation of the $^{12}$C(AMD)+$\alpha$
may not be sufficient for the $1p$-$1h$ excitation because an $\alpha$ cluster around 
$^{12}$C is assumed in the model though twelve nucleon dynamics is incorporated in the
wave function of the $^{12}$C AMD wave functions.
Instead of cluster model calculations, mean-field calculations including particle-hole excitations such as the random phase approximation (RPA) have been applied to investigate ISGMR. In the RPA calculations for 
$^{16}$O \cite{Blaizot:1976cw,Ma:1997zzb,Papakonstantinou:2009zz,Gambacurta:2010nk}, 
it was found that monopole strength spreads out and has a multi peak structure 
with the centroid around $E_x=20 \sim 25$ MeV. They describe the experimental 
strength in the high-energy region 
$E_x \ge 16$ MeV measured by $(\alpha,\alpha')$ scattering. 
However, the peak structure with the significant fraction of EWSR in the low-energy part 
are not reproduced by the mean-field calculations. 

To take into account the coexistence of cluster and mean-field features 
in the ISM excitation, we extend our present framework of the $^{12}$C(AMD)+$\alpha$GCM
by adding the $1p$-$1h$ type basis wave function on the top of the approximate ground state
wave function obtained by the $^{16}$O(AMD+VAP) calculation. 
After explaining the additional basis wave functions, we discuss the monopole transition 
in $^{16}$O.

\subsection{AMD+VAP calculation of $^{16}$O and $1p$-$1h$ excitation}

The present method of the $^{12}$C(AMD)+$\alpha$GCM is suitable mainly 
to describe various types of cluster excitation. To take into account the $1p$-$1h$ excitation, we perform the AMD+VAP calculation for $^{16}$O and consider 
small variations of single-particle wave functions from the obtained ground state wave function.
In a similar way to Eq.~\ref{eq:AMD} for $^{12}$C, 
an AMD wave function for $^{16}$O is written by 
a Slater determinant of 16 single-nucleon Gaussian wave packets, 
\begin{equation}\label{eq:AMD-16O}
\Phi_{^{16}{\rm O}}^{\rm AMD}({\bf Z})=\frac{1}{\sqrt{A!}} {\cal{A}} \{
  \varphi_1,\varphi_2,...,\varphi_{A} \}.
\end{equation}
In the AMD+VAP method, the energy variation is done with respect to the spin-parity
eigen wave function $P^{J\pi}_{MK}\Phi^{\rm AMD}_{^{16}{\rm O}}({\bf Z})$. 
After the AMD+VAP calculation for $^{16}$O, we get the optimum AMD solution 
$\Phi^{\rm AMD}_{^{16}{\rm O}}({\bf Z}_0)$
which is approximately regarded as the intrinsic wave function of the ground state.
Here ${\bf Z}_0$ indicates the set of optimized parameters 
${\bf Z}_0=\{{\bf X}^0_1,{\bf X}^0_2,\cdots,{\bf X}^0_A, \xi_1,\cdots,\xi_A \}$.

Then, we vary the spatial part of each single-particle wave function from 
the AMD wave function, $\Phi^{\rm AMD}_{^{16}{\rm O}}({\bf Z}_0)$, by shifting 
a Gaussian center of the $i$th single-particle wave function,
${\bf X}^0_i\rightarrow {\bf X}^0_i+\delta {\bf e}_\sigma$ $(\sigma=1,2,3)$.
(${\bf e}_1$, ${\bf e}_2$ and ${\bf e}_3$ are the three-dimension unit vectors.)
For all single-particle wave function, we consider a small shift to 
three directions independently, namely, $A\times 3$ kinds of shifted 
wave functions $\Phi^{\rm AMD}_{^{16}{\rm O}}({\bf Z}'_0(i,\sigma))$ 
($i=1,\cdots,A$ and $\sigma=1,2,3$) with
${\bf Z}'_0(i,\sigma)\equiv 
\{{\bf X}^0_1,\cdots,{\bf X}^0_i+\delta {\bf e}_\sigma,\cdots,
{\bf X}^0_A,\xi_1,\cdots,\xi_A \}$.
By using the linear combination of 47 wave functions,
the original wave function $\Phi^{\rm AMD}_{^{16}{\rm O}}({\bf Z}_0)$ 
and the shifted ones $\Phi^{\rm AMD}_{^{16}{\rm O}}({\bf Z}'_0(i,\sigma))$, 
$1p$-$1h$ excitations in the intrinsic frame are incorporated. 
We fix the spin orientations $\xi_i$ and
consider the $1p$-$1h$ excitations mainly for spatial part.
For excited $0^+$ states of $^{16}$O, we superpose 
the spin-parity eigen states projected from 
those wave functions, $P^{J^\pi}_{MK}\Phi^{\rm AMD}_{^{16}{\rm O}}({\bf Z}_0)$
and $P^{J^\pi}_{MK}\Phi^{\rm AMD}_{^{16}{\rm O}}({\bf Z}'_0(i,\sigma))$.
The coefficients of each basis wave functions are determined by diagonalizing the norm and 
Hamiltonian matrices.
We call this calculation "$^{16}$O(AMD)+$1p$-$1h$".

In addition to the $^{16}$O(AMD)+$1p$-$1h$ calculation 
in the $1p$-$1h$ model space, we also perform the hybrid calculation of 
$^{12}$C(AMD)+$\alpha$GCM and $^{16}$O(AMD)+$1p$-$1h$ by superposing 
all basis wave functions. The coefficients are determined again by the 
diagonalization. 

\subsection{Monopole transitions}
The strength function of the ISM excitation from
the ground state of $^{16}$O is 
\begin{eqnarray}
&&S(E)\equiv \delta(E-E_n) |M(IS0,0^+_1\rightarrow 0^+_n)|^2 \\
&&M(IS0,0^+_1\rightarrow 0^+_n)=\langle 0^+_n|\sum^{16}
_{i=1} {\bf r}_i^2|0^+_1 \rangle.
\end{eqnarray}
For the isoscalar excitation, this is 4 times as much as 
the isoscalar $E0$ strength function defined in Refs.~\cite{Krison76,Lui:2001xh}.
The EWSR of the ISM transition is 
\begin{equation}
\sum_n(E_n-E_1)|M(IS0,0^+_1\rightarrow 0^+_n)|^2=\frac{2\hbar^2}{m}16\langle r^2\rangle,
\end{equation}
where $\langle r^2\rangle$ is the mean square matter radius of the ground state, 
\begin{equation}
\langle r^2\rangle=\frac{1}{16}\langle 0^+_1|\sum_{i=1}^{16} {\bf r}_i^2 |0^+_1\rangle.
\end{equation}
In the results of the $^{12}$C(AMD)+$\alpha$GCM, the $^{16}$O(AMD)+$1p$-$1h$, and the hybrid of 
$^{12}$C(AMD)+$\alpha$GCM and $^{16}$O(AMD)+$1p$-$1h$, 
the energy weighted sum of the ISM strength for all excited states 
is 93\%, 87\%, and 95\% of the EWSR value,
and that for excited states up to 40 MeV ($E_x\le 40$ MeV) is 77\%, 64\%, 69\%, respectively. 

\begin{figure}[th]
\epsfxsize=0.35\textwidth
\centerline{\epsffile{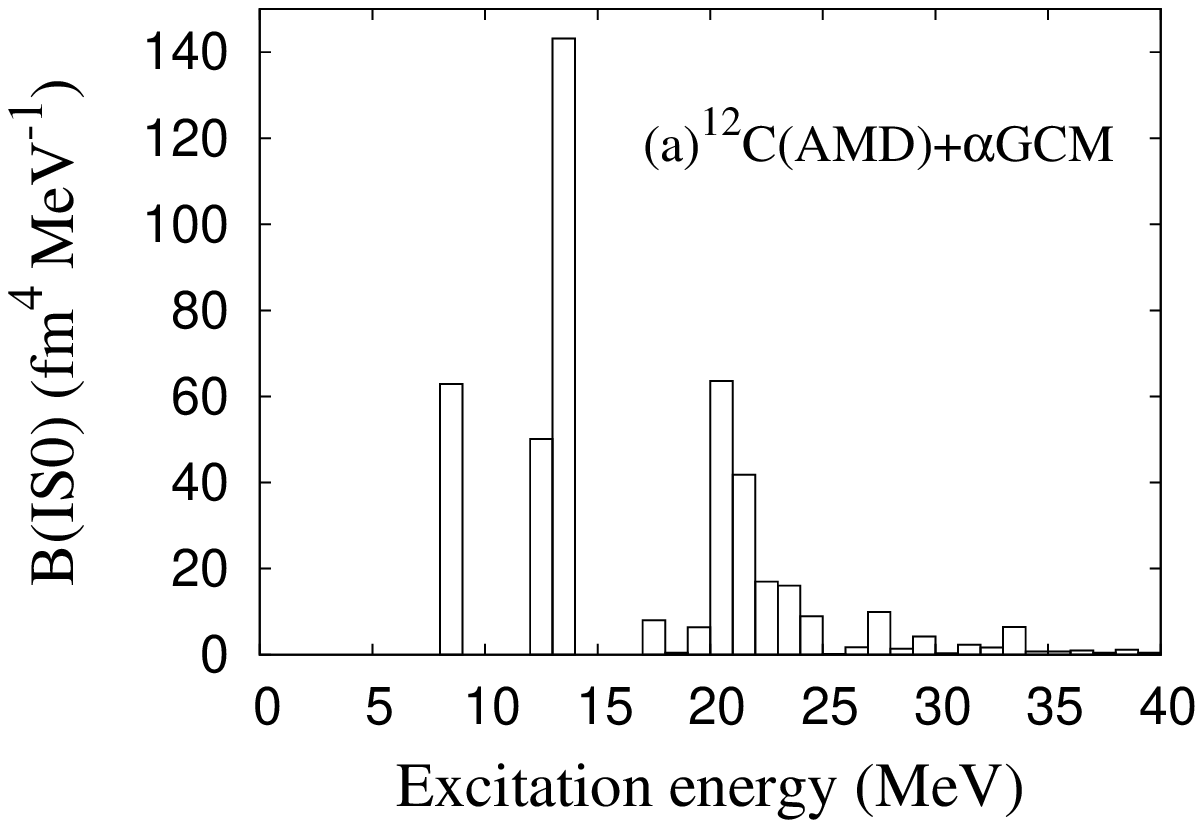}}
\epsfxsize=0.35\textwidth
\centerline{\epsffile{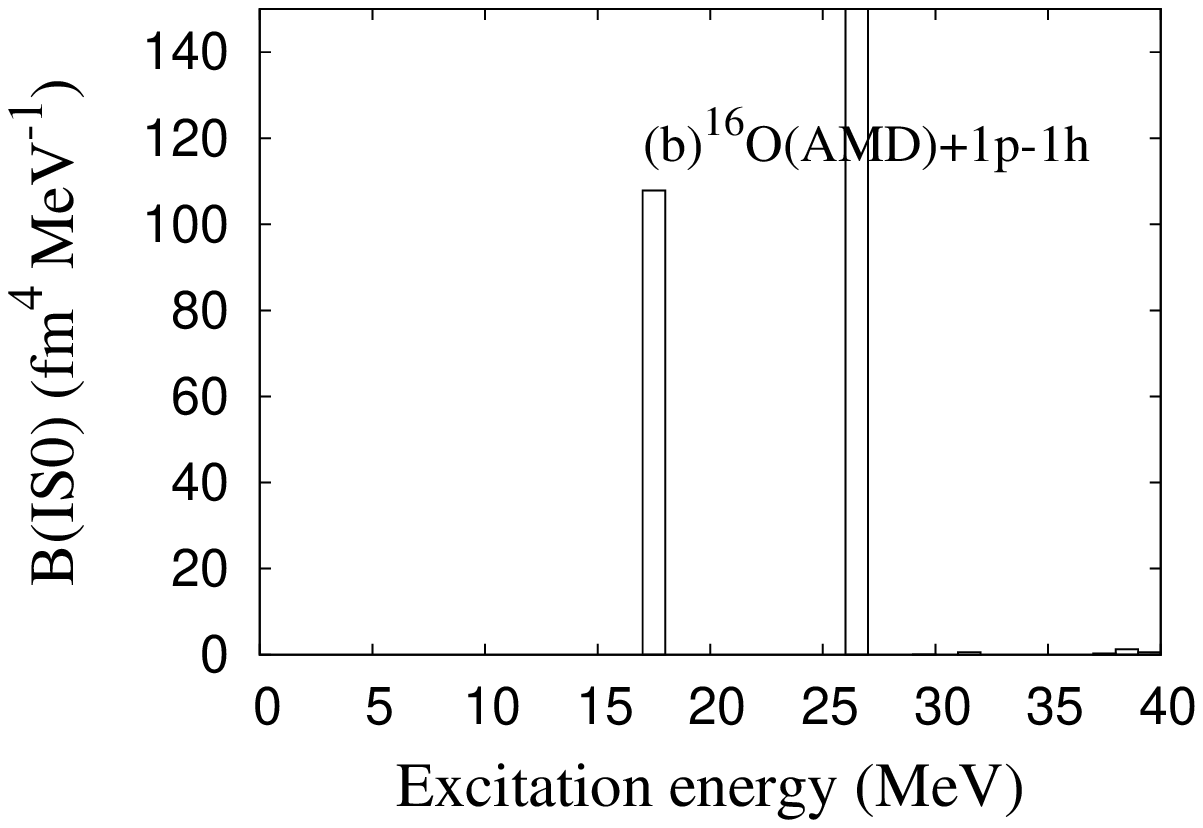}}
\epsfxsize=0.35\textwidth
\centerline{\epsffile{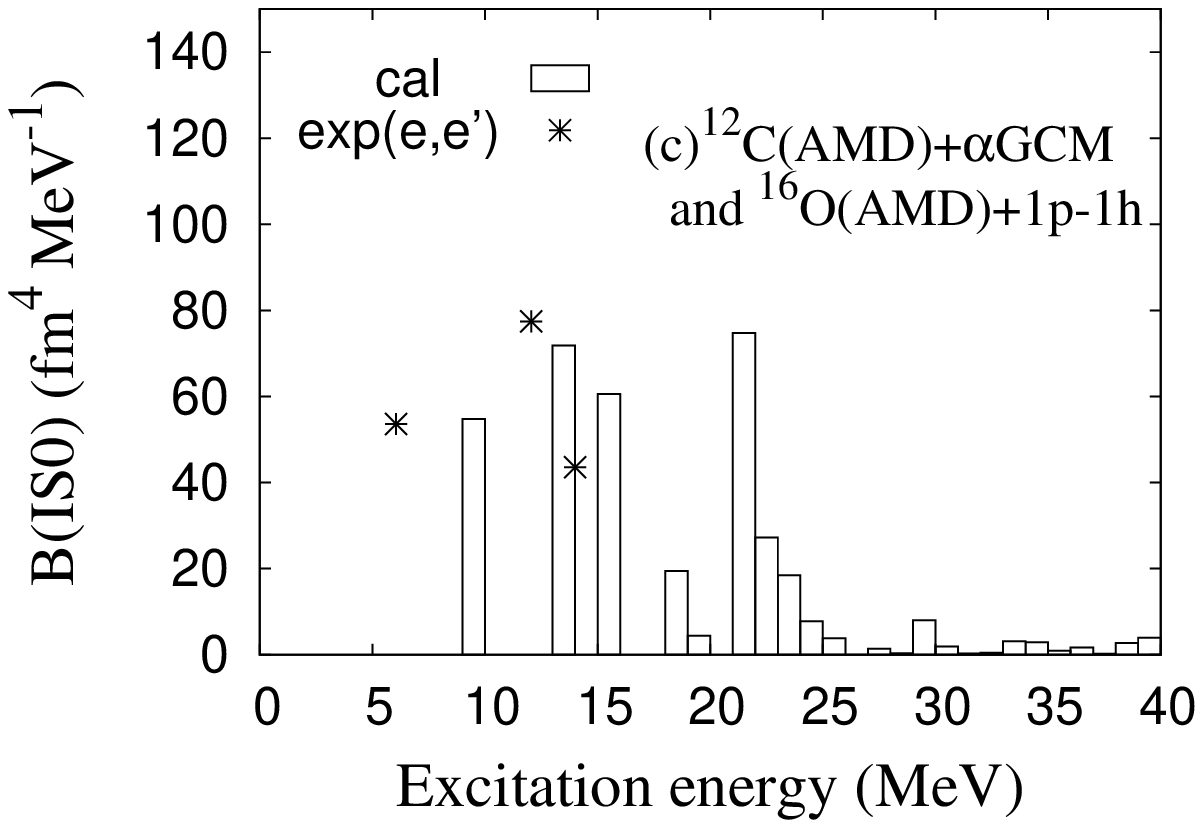}}
\epsfxsize=0.35\textwidth
\centerline{\epsffile{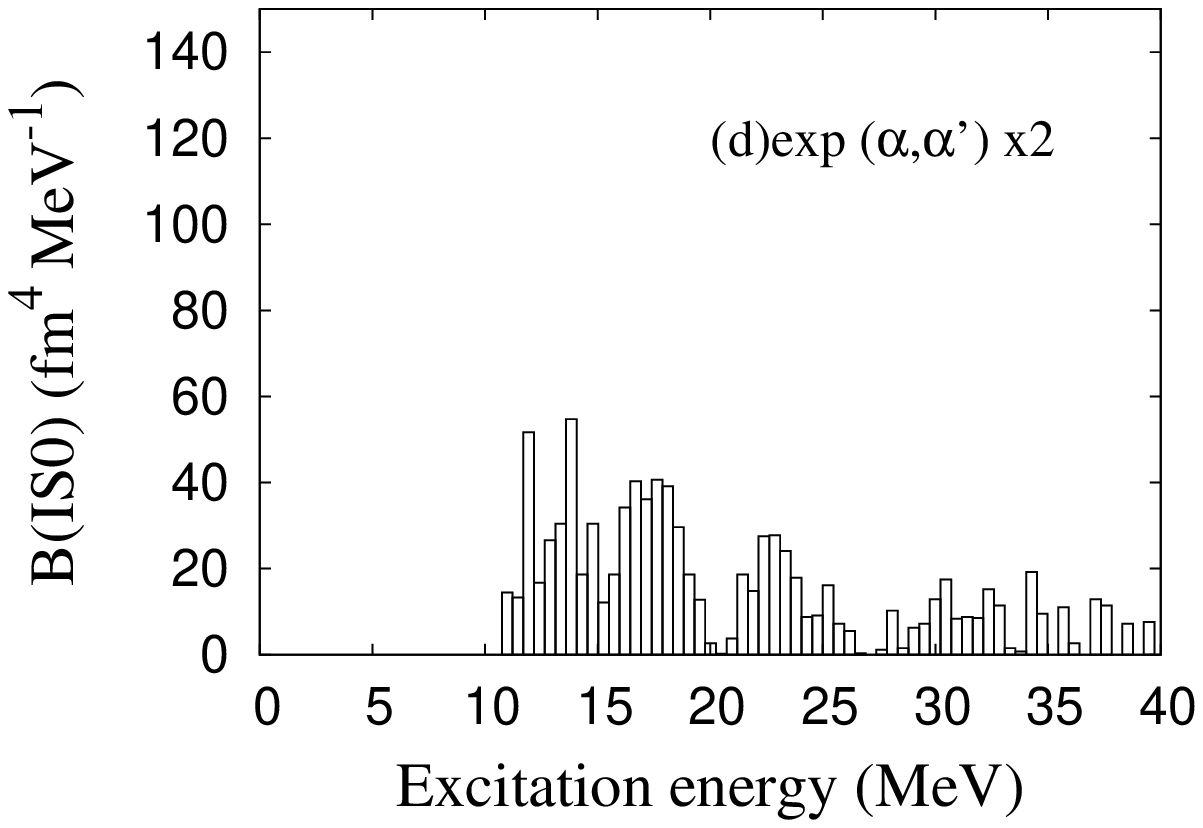}}
\caption{
\label{fig:is0}
Isoscalar monopole transition strength function.
The theoretical $B(IS0)$ calculated with (a)the 
$^{12}$C(AMD)+$\alpha$GCM, (b)the $^{16}$O(AMD)+$1p$-$1h$, and (c)the hybrid of 
$^{12}$C(AMD)+$\alpha$GCM and $^{16}$O(AMD)+$1p$-$1h$. 
In the histogram
the strength in each energy bin is summed up.
The experimental $B(IS0)$ (fm$^4$) 
converted from the $B(E0)$ measured by $(e,e')$ scattering 
for the $0^+$ states at 6.05 MeV, 12.05 MeV, and 14.01 MeV 
are also shown by stars in the third panel(c).
(d)The experimental data measured by $(\alpha,\alpha')$ scattering.
We multiply the data from Ref.~\cite{Lui:2001xh} by a factor 2 in the panel (d).
}
\end{figure}

In the present calculation, all excited states are discrete states because 
of the bound state approximation. We calculate 
the ISM transition matrix element $M(IS0)$ for $0^+_n$ states of the 
$^{12}$C(AMD)+$\alpha$GCM, the $^{16}$O(AMD)+$1p$-$1h$, and the hybrid full calculations.
The calculated ISM transition strength ($B(IS0)=|M(IS0)|^2$) is shown in the histogram 
in Fig.~\ref{fig:is0}, 
where the strength $|M(IS0)|^2$ for $0^+_n$ states in each energy bin 
is summed up. 

\begin{figure}[th]
\epsfxsize=0.35\textwidth
\centerline{\epsffile{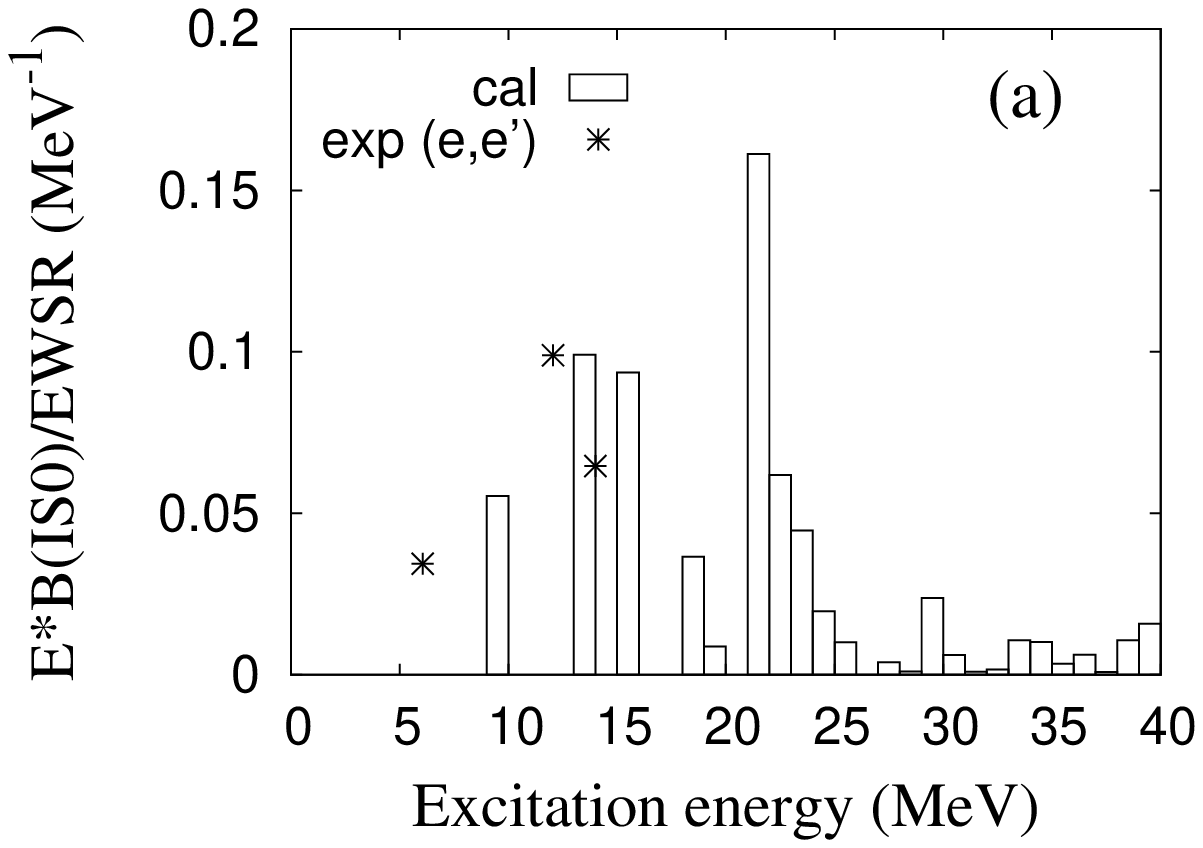}}
\epsfxsize=0.35\textwidth
\centerline{\epsffile{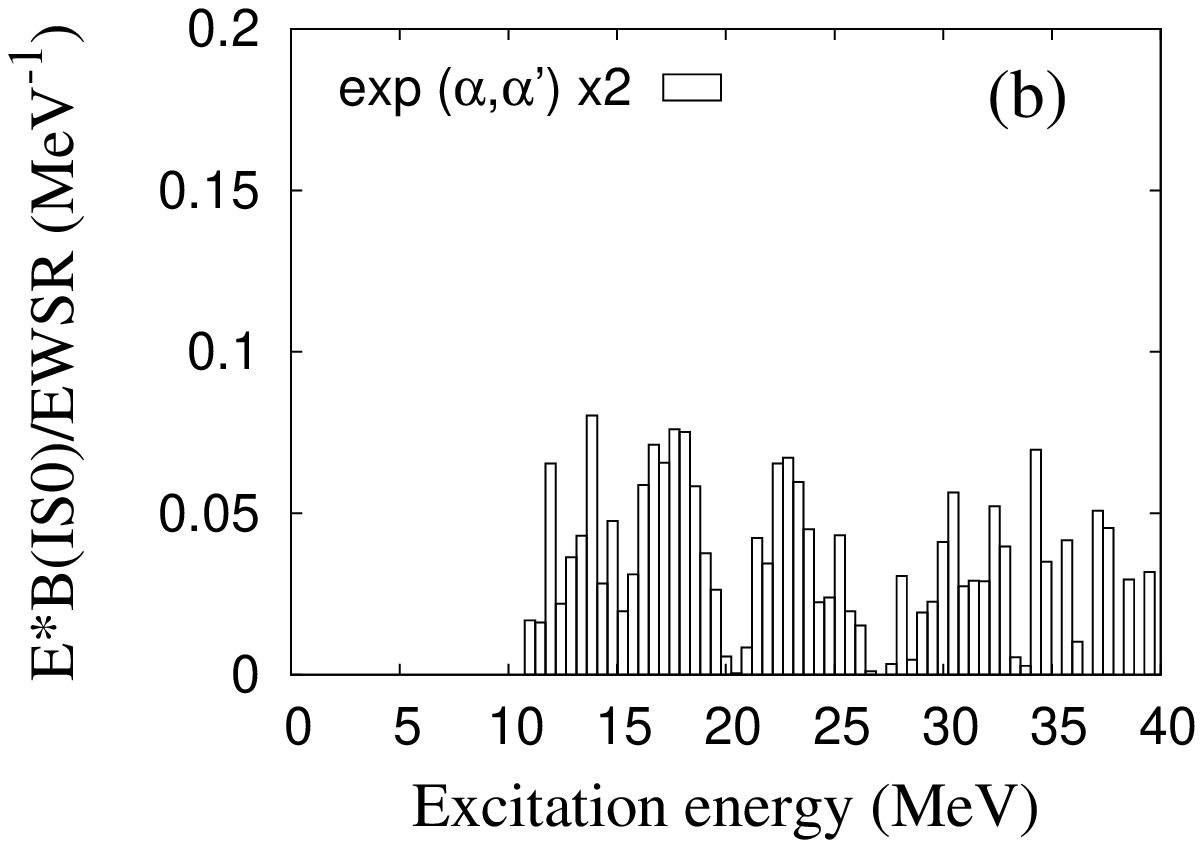}}
\caption{
\label{fig:ewsr}
The ratio of the energy weighted sum to the total EWSR.
(a)The theoretical values calculated with the 
the hybrid of 
$^{12}$C(AMD)+$\alpha$GCM and $^{16}$O(AMD)+$1p$-$1h$.
(b)The experimental data measured by $(\alpha,\alpha')$ scattering.
We multiply the data from Ref.~\cite{Lui:2001xh} by a factor 2 in the plot.
}
\end{figure}

In the $^{12}$C(AMD)+$\alpha$GCM result, the significant strength exists in the low-energy part 
for the $0^+_{II}$, $0^+_{III}$, and $0^+_{IV}$ states having 
the $^{12}$C($0^+_1$,$2^+_1$)+$\alpha$ cluster structures. They exhaust 
$\sim 30\%$ of the EWSR. Such the large fraction in the low-energy part ($E_x\le 16$ MeV)
is comparable to the $4\alpha$-OCM calculation where $\sim 20\%$ of the EWSR
exists in the $E_x\le 16$ MeV part \cite{Yamada:2011ri}.
In higher-energy region, 
the strength concentrates around the region  $E_x\sim 20$ MeV.
The EWSR ratio of the high-energy part ($16\le E_x \le 40$ MeV) is $\sim 45\%$ 
in the $^{12}$C(AMD)+$\alpha$GCM calculation.

In the result of the $^{16}$O(AMD)+$1p$-$1h$ calculation, the ISM transition strength shows the two-peak structure around $E_x\sim$ 20 MeV, 
one below and the other above $E_x=20$ MeV.
The higher peak corresponds to the breathing mode which can be described by the coherent 
isotropic single-particle motion, while  
the lower peak is understood as the motion of one $\alpha$-cluster against the 
$^{12}$C core. The latter mode originates in the ground state 
$\alpha$ correlation around the $^{12}$C core which is contained 
in the AMD+VAP result of $^{16}$O($0^+_1$). 
The lower and the higher peaks exhaust about 20\% and 40\% of the EWSR, 
respectively. The EWSR ratio for the lower peak is the same order of the EWSR ratio for 
the cluster states with the $^{12}$C($0^+_1$,$2^+_1$)+$\alpha$ cluster structures
in $E_x \le 16$ MeV calculated with the $^{12}$C(AMD)+$\alpha$GCM.

The full calculation using the hybrid model space of the $^{12}$C(AMD)+$\alpha$GCM
and the $^{16}$O(AMD)+$1p$-$1h$ shows qualitatively similar features of the 
$^{12}$C(AMD)+$\alpha$GCM calculation. Namely, there exist 
three peaks corresponding to the cluster states in the low-energy part ($E_x\le 16$ MeV), 
and the concentration of the strength around the peak-like structure slightly above 20 MeV.  
The EWSR ratios of the low-energy part ($E_x\le 16$ MeV) and the high-energy part
($16\le E_x\le 40$ MeV) are $\sim 25\%$ and $\sim 40\%$, respectively. 

Comparing the results of the $^{12}$C(AMD)+$\alpha$GCM, the $^{16}$O+($1p$-$1h$), and the full hybrid calculations, 
it is found that there is no significant difference of 
the EWSR ratios of the low-energy and high-energy parts among three calculations.
It implies that two modes around $\sim$ 20 MeV obtained in the $^{16}$O(AMD)+$1p$-$1h$ are 
involved in excited states of the $^{12}$C(AMD)+$\alpha$GCM.
That is, the higher peak of the collective breathing mode corresponds to 
the peak-like structure slightly above $20$ MeV in the $^{12}$C(AMD)+$\alpha$GCM and
the full calculation, while the lower mode for the $^{12}$C-$\alpha$ motion 
is fragmented in the lowest three excited $0^+$ states with the 
$^{12}$C($0^+_1$,$2^+_1$)+$\alpha$ cluster structures.
Namely, we can conclude the origins of isoscalar monopole excitations as follows.
In the mean-field type $1p$-$1h$ excitation there exist two modes around $E_x\sim 20$ MeV.
The lower mode corresponds to the $^{12}$C-$\alpha$ relative motion and the higher one 
is the collective breathing mode. Because of the coupling with the cluster excitation, 
the lower mode is fragmented into several cluster states in $E_x \le 16$ MeV
while lowering the energy centroid. The strength of the higher breathing mode 
is somehow spread and also its energy centroid is lowered  to 
contribute to the strength around $E_x \sim 20$MeV.

The ISM transition strength has been observed by $(\alpha,\alpha')$ scattering \cite{Lui:2001xh}.
The measured strength for the $0^+$ states at 12 and 14 MeV is
smaller than the that observed by 
$(e,e')$ scattering by a factor $2-4$. Moreover, their measurement in the energy region 
$11 <E_x <40$ MeV
yields only $\sim$50$\%$ of the $E0$ EWSR. These fact may suggest possible
ambiguity of the normalization in the ISM strength measured by
$(\alpha,\alpha')$  scattering. We multiply the experimental data by a factor 2 and 
show the values in Fig.~\ref{fig:is0}(d) to compare the shape of strength function 
with our result. 
Comparing the result of the full calculation with the 
experimental data, it is shown that the strength for the $0^+_{III}$ and
$0^+_{IV}$ states at 13 and 15 MeV may describe the peaks 
in the $11 < E_x < 16$ MeV of the experimental data.
The significant strength in the higher region around 20 MeV is considered to
correspond to the bump structures in the regions 
$16 < E_x <20$ MeV and/or $20 < E_x <25$ MeV. The calculated strength are not fragmented so much
as the experimentally measured one, maybe, because of the limitation of the present 
model space. 
The EWSR ratio of the full calculation and 
that of the experimental data are shown in Fig.~\ref{fig:ewsr}.
We again multiply the experimental data of Ref.~\cite{Lui:2001xh} measured by $(\alpha,\alpha')$ scattering by a factor 2 in the plotting.

\section{Summary and outlooks}\label{sec:summary}
Cluster structures and monopole transitions in positive parity states of $^{16}$O were 
investigated based on the $^{12}$C(AMD)+$\alpha$GCM calculation.
The lowest three excited $0^+$ states ($0^+_{II}$,  $0^+_{III}$, and $0^+_{IV}$) have
the $^{12}$C($0^+_1$,$2^+_1$)+$\alpha$ cluster structures. 
The $0^+_{II}$ with the $^{12}$C($0^+_1$)+$\alpha$ structure and its rotational
band members qualitatively reproduce the properties such as energy levels and $E2$ and monopole
transition strengths for the experimental 
$0^+_2$, $2^+_1$, and $4^+_1$ states, which have been considered to be the  
 $^{12}$C($0^+_1$)+$\alpha$ cluster band. As far as we know, the present calculation is the 
first microscopic calculation that can describe reasonably the excitation energies 
of these excited states.

In the present calculation, we obtained the fifth $0^+$ state $(0^+_V)$ having the developed 
$^{12}$C($0^+_2$)+$\alpha$ structure. 
Because of the feature that an $\alpha$ cluster is moving in the $L=0$ wave 
around the $^{12}$C($0^+_2$),
it is regarded as the
$4\alpha$ cluster gas state similar to the 3$\alpha$ cluster gas in the $^{12}$C($0^+_2$). 
This state may correspond to the $0^+_6$ state of the $4\alpha$ cluster gas state suggested in the 
$4\alpha$-OCM by Funaki {\it et al.} \cite{Funaki:2008gb,Funaki:2010px}. 

With the analyses of the $E2$ transition strength and the $^{12}$C($0^+_2$)+$\alpha$ component, 
we considered band members of the $^{12}$C($0^+_2$)+$\alpha$ cluster state. 
Around $E_x\sim 20$ MeV, there are several $2^+$ and $4^+$ states 
having some component of $^{12}$C($0^+_2$)+$\alpha$. The $E2$ transition 
strength is fragmented among them. The present result suggests that  the  
structure change, in other words, the state mixing occurs in the rotation 
of the $^{12}$C($0^+_2$)+$\alpha$ cluster structure. 
It make it difficult to assign clearly the $^{12}$C($0^+_2$)+$\alpha$ band members in high spin states. 
This feature is different from that of the $^{12}$C($0^+_1$)+$\alpha$ cluster band 
and may originate in the 3$\alpha$ cluster gas feature of 
the $^{12}$C($0^+_2$) that 
might be fragile in the rotation. 

The isoscalar monopole excitation was discussed with the $^{12}$C(AMD)+$\alpha$GCM 
and also with the hybrid calculation of the $^{12}$C(AMD)+$\alpha$GCM  and 
$^{16}$O(AMD)+$1p$-$1h$. 
In the strength of both calculations, 
there exist three peaks for the cluster
states in the low-energy part ($E_x < 16$ MeV). This is consistent with the 
preceding work with the $4\alpha$-OCM calculation \cite{Yamada:2011ri}.
We also found the concentration 
of the strength around the peak-like structure slightly above $E_x \sim 20$ MeV, which  
originates in the collective breathing mode. 
Comparing the hybrid calculation with the $^{16}$O+$1p$-$1h$ calculation, 
we conclude the origins of isoscalar monopole excitations as follows.
In the mean-field type $1p$-$1h$ excitation there exist two modes around $E_x=20$ MeV.
The lower mode corresponds to the $^{12}$C-$\alpha$ relative motion and the higher one 
is the collective breathing mode. Because of the coupling with the cluster excitation, 
the lower mode is fragmented into several cluster states in $E_x < 16$ MeV
while lowering the energy centroid. 
The higher-energy breathing mode 
is somehow spread and its energy centroid is lowered  to 
contribute to the strength around $E_x \sim 20$ MeV.

The present calculation is a bound state approximation. The stability of the 
excited states should be studied in more details by taking into account coupling with 
continuum states. We also should reexamine the choice of the effective interaction and the 
interaction parameters for quantitative reproduction of energy levels.
In the present work, we used the same phenomenological 
effective nuclear forces as those used in the 
previous work on $^{12}$C. The energy spectra of $^{16}$O 
may be improved by fine tuning of the interaction parameters.
However, we have some difficulty in completely reproducing
the binding energies of $\alpha$, $^{12}$C, and $^{16}$O 
as well as the energy spectra of the subsystem $^{12}$C simultaneously 
with such the phenomenological effective nuclear interaction. 
{\it Ab initio} calculation
based on realistic nuclear force is one of the promising tools  
for quantitative prediction of energy spectra of $^{16}$O 
though applications of 
{\it ab initio} calculations to cluster states are still limited.

\section*{Acknowledgments} 
The author would like to thank Dr.~Funaki and Dr.~Yamada for fruitful discussions.
The computational calculations of this work were performed by using the
supercomputers at YITP.
This work was supported by Grant-in-Aid for Scientific Research from Japan Society for the Promotion of Science (JSPS).
It was also supported by 
the Grant-in-Aid for the Global COE Program "The Next Generation of Physics, 
Spun from Universality and Emergence" from the Ministry of Education, Culture, Sports, Science and Technology (MEXT) of Japan. 


\end{document}